\documentclass[acmsmall,screen=true,anonymous=false,bookmarks=true]{acmart}
\usepackage{amsmath,amsfonts}
\usepackage{multirow}
\usepackage{multicol}
\usepackage{amsthm}
\usepackage{listings}
\usepackage{algorithm}
\usepackage{algpseudocode}
\usepackage{graphicx}
\usepackage{textcomp}
\usepackage{xcolor}
\usepackage{cleveref}
\usepackage{booktabs}
\usepackage{comment}
\usepackage{tabularx}
\usepackage{multirow}
\usepackage{bm}
\usepackage{url}
\usepackage{xspace}
\usepackage{pifont}
\usepackage{arydshln}
\usepackage{verbatim}
\usepackage[referable]{threeparttablex}
\usepackage{calc} 
\usepackage{soul}
\usepackage{color}
\usepackage{float}
\usepackage{subcaption}
\usepackage[utf8]{inputenc}
\usepackage{booktabs} 
\usepackage{tabularx} 
\usepackage{tikz}
\usetikzlibrary{shapes.geometric, arrows}

\tikzstyle{startstop} = [rectangle, rounded corners, minimum width=3cm, minimum height=1cm,text centered, draw=black]
\tikzstyle{io} = [trapezium, trapezium left angle=70, trapezium right angle=110, minimum width=3cm, minimum height=1cm, text centered, draw=black]
\tikzstyle{process} = [rectangle, minimum width=3cm, minimum height=1cm, text centered, draw=black]
\tikzstyle{arrow} = [thick,->,>=stealth]

\usepackage{filecontents}                                  
\usepackage{pgfplots}
\usepackage{pgfplotstable}
\usepackage{scalefnt}
\pgfplotsset{compat=newest}
\usepgfplotslibrary{groupplots}

\renewcommand{\vec}[1]{\boldsymbol{#1}}    

\definecolor{USTgold}{RGB}{153,102,0}
\definecolor{USTyellow}{RGB}{204,153,0}
\definecolor{USTyellowlight}{RGB}{255,212,0}
\definecolor{USTorange}{RGB}{255,166,26}
\definecolor{USTpink}{RGB}{255,157,157}
\definecolor{USTblue}{RGB}{0,51,102}
\definecolor{USTmiddle}{RGB}{0,116,188}
\definecolor{USTlight}{RGB}{99,202,225}
\definecolor{USTgray}{RGB}{204,204,204}
\definecolor{USTred}{RGB}{237,27,47}
\definecolor{USTdarkred}{RGB}{124,35,72}

\definecolor{CUHKorange}{RGB}{244,106,18} 
\definecolor{CUHKblue}{RGB}{0,111,190}    
\definecolor{CUHKgreen}{RGB}{0,127,128}   
\definecolor{CUHKred}{RGB}{228,46,36}     
\definecolor{CUHKyellow}{RGB}{198,148,34} 
\definecolor{CUHKdark}{RGB}{114,44,114}   
\definecolor{CUHKmiddle}{RGB}{144,44,144} 
\definecolor{CUHKlight}{RGB}{167,44,167}

\iftrue
\def\BibTeX{{\rm B\kern-.05em{\sc i\kern-.025em b}\kern-.08em
    T\kern-.1667em\lower.7ex\hbox{E}\kern-.125emX}}

\fi

\usepackage{tcolorbox}
\tcbuselibrary{skins,breakable}
    {\endtcolorbox}

\RequirePackage[normalem]{ulem} 
\RequirePackage{color}\definecolor{RED}{rgb}{1,0,0}\definecolor{BLUE}{rgb}{0,0,1} 
\providecommand{\DIFadd}[1]{{\protect\color{blue}{#1}}}                           

\newcommand{\todo}[1]{\textcolor{red}{[TODO: #1]}}
\newcommand{\revise}[1]{\DIFadd{#1}}

\usepackage{xcolor,graphicx}
\usepackage{ulem}

\setcopyright{acmcopyright}
\acmJournal{TODAES}
\graphicspath{{./fig/}}

\DeclareMathAlphabet\mathbfcal{OMS}{cmsy}{b}{n}
\newcommand{\ten}[1]{\mathbfcal{#1}}

\definecolor{myblue}{RGB}{73,148,196}   
\definecolor{mydarkblue}{RGB}{18,38,79} 
\definecolor{myorange}{RGB}{234,85,20}  
\definecolor{myyellow}{RGB}{250,192,61} 
\definecolor{mypink}{RGB}{252,228,215}  
\definecolor{mygreen}{RGB}{19,138,7}  

\usepackage{empheq}

\begin{document}

\setcounter{page}{1}

\title{RL-MUL 2.0: Multiplier Design Optimization with Parallel Deep Reinforcement Learning and Space Reduction
}

\author{Dongsheng Zuo}
\affiliation{
    \institution{The Hong Kong University of Science and Technology (Guangzhou)}
}
\author{Jiadong Zhu}
\affiliation{
    \institution{The Hong Kong University of Science and Technology (Guangzhou)}
}
\author{Yikang Ouyang}
\affiliation{
    \institution{The Hong Kong University of Science and Technology (Guangzhou)}
}
\author{Yuzhe Ma}
\affiliation{
    \institution{The Hong Kong University of Science and Technology (Guangzhou)}
}

\thanks{Dongsheng Zuo and Jiadong Zhu contributed equally to this research.}
\thanks{This work is supported in part by the Department of Education of Guangdong Province (No. 2024KTSCX037), the Guangzhou-HKUST(GZ) Joint Funding Program (No. 2023A03J0155), and the Guangzhou Municipal Science and Technology Project (Municipal Key Laboratory Construction Project, Grant No.2023A03J0013).}

\begin{abstract}
Multiplication is a fundamental operation in many applications, and multipliers are widely adopted in various circuits. 
However, optimizing multipliers is challenging due to the extensive design space. 
In this paper, we propose a multiplier design optimization framework based on reinforcement learning.
We utilize matrix and tensor representations for the compressor tree of a multiplier, enabling seamless integration of convolutional neural networks as the agent network. 
The agent optimizes the multiplier structure using a Pareto-driven reward customized to balance area and delay. 
Furthermore, we enhance the original framework with parallel reinforcement learning and design space pruning techniques and extend its capability to optimize fused multiply-accumulate (MAC) designs. 
Experiments conducted on different bit widths of multipliers demonstrate that multipliers produced by our approach outperform all baseline designs in terms of area, power, and delay. 
The performance gain is further validated by comparing the area, power, and delay of processing element arrays using multipliers from our approach and baseline approaches.
\end{abstract}

\maketitle

\pagestyle{plain}

\section{Introduction}
\label{sec:intro}
In the era of rapid advancements in neural networks and streaming media applications, the demand for computational power has intensified.
Notably, the multiply-accumulate (MAC) computation can constitute over 99\% of operations in standard deep neural networks.
At the hardware layer, multipliers and MAC circuits are integral to the architecture of compute-intensive circuits, 
significantly impacting the system performance, energy consumption, spatial requirements, and design complexities. 
Therefore, swiftly designing multipliers and MACs that meet metric specifications such as power, performance, and area (PPA) becomes imperative.


%
%

Multiplier design optimization at the architecture level is non-trivial due to the huge design space.
For an 8-bit multiplier, the design space size is on the order of \(10^9\), while for a 16-bit multiplier, it reaches approximately \(10^{23}\).
This exponential scaling highlights the significant complexity involved in effectively exploring and optimizing designs as bit-width increases.
The multiplier design is fundamentally segmented into three primary components: a partial product generator (PPG), a compressor tree (CT), and a carry propagation adder (CPA). 
Among these, the optimization of the Compressor Tree (CT) is pivotal, as it significantly influences the PPA of a multiplier. 
The architecture of the compressor tree was first introduced in \cite{Datapath-1964TC-Wallace}, designed for parallel compression (i.e., addition) of partial products in multiplication operations. 
This innovation has enabled the application of compressor trees in other datapath circuits, such as MACs and vector adders. 
Conventionally, MAC operations extend the functionality of multipliers by incorporating an accumulator after multiplication, resulting in increased operational delay. 
In contrast, the merged MAC is proposed, which enables the execution of MAC operations within the multiplication time by integrating the addend directly into the partial products~\cite{Datapath-1997ARITH-Stelling}. 
This approach also allows the optimization methodologies developed for multipliers to be applied to MAC design optimization.
Generally, datapath designs, including adders, multipliers, and MACs, can be completed manually.
Take multiplier design as an example. 
The manual design includes Wallace tree structure~\cite{Datapath-1964TC-Wallace}, Dadda tree structure \cite{Datapath-1983ARITH-Dadda}, and further optimized designs based on them~\cite{Datapath-1993TVLSI-Fadavi-Ardekani,Datapath-2005ISCAS-Itoh,Datapath-1993ASAP-Bickerstaff,Datapath-2014ATC-Luu}, which effectively optimize area, power, and performance for specific technology nodes and applications. 
The Wallace tree strategically organized the compressor layers \cite{Datapath-1993TVLSI-Fadavi-Ardekani}. 
An area-reduced tree is proposed by using a maximum number of 3:2 compressors early and carefully placing 2:2 compressors \cite{Datapath-1993ASAP-Bickerstaff}.
Itoh \textit{et al.}~\cite{Datapath-2005ISCAS-Itoh} proposed an advanced rectangular-styled tree structure, tailored specifically for $32\textrm{-bit}\times24\textrm{-bit}$ multipliers. 
Optimizations for merged MAC structures have been explored based on the characteristics of multiply-accumulate operations. 
Basiri \textit{et al.} \cite{Datapath-2014TODAES-Basiri} proposed a high-radix Booth-encoded merged MAC targeting floating-point DSP applications. 
Their design combines Wallace and Braun tree structures to optimize circuit depth and area, effectively balancing performance and resource usage for floating-point operations.
Tung \textit{et al.} \cite{Datapath-2020IEEEAccess-Tung} proposed a method where the final addition and accumulation of higher significance bits are merged to the partial products of the next multiplication operation.
Zhang \textit{et al.} \cite{Datapath-2021ASPDAC-Zhang} proposed a strategy optimizing pipeline merged MAC.
These regular structure-based designs may not always meet the stringent PPA specifications required.
To address this, full custom-designed multipliers are developed, which are finely optimized for specific fabrication processes or unique application scenarios \cite{Datapath-2020SSCL-Shavit,Datapath-2022MysuruCon-Jangalwa,Datapath-2023ICSPC-R}. 
However, a significant engineering effort is required to explore the huge design space with manual design, which limits design flexibility and efficiency.

The automatic generation or search methods have provided a more flexible solution to datapath designs.
A three-dimensional method for designing the compressor tree was proposed, which utilized an input-to-output delay model \cite{Datapath-1996TC-Oklobdzija,Datapath-1995ARITH-Martel,Datapath-1998TC-Stelling}.
Integer linear programming (ILP) is another widely investigated approach for datapath circuit optimization.
Xiao~\textit{et al.} \cite{Datapath-2021DATE-Xiao} employed ILP for global optimization of multiplier design by minimizing the total number of compressors in the compressor tree.
In addition, ILP has also been applied for exploring adder trees based on analytical area, power, and timing models \cite{Datapath-2007ASPDAC-Liu}.
However, these works may suffer from the long runtime of the ILP solver as well as the misaligned objective between the modeled PPA metrics and real synthesized metrics.
Heuristic search strategies utilize various pruning techniques and avoid exhaustive searches \cite{Datapath-2003ICCAD-Liu, Datapath-2013DAC-Roy, Datapath-2008ASPDAC-Parandeh-Afshar, Datapath-2011TRETS-Parandeh-Afshar, Datapath-2018TC-Kumm}.
A heuristic is introduced in \cite{Datapath-2008ASPDAC-Parandeh-Afshar} for the design of compressor trees using generalized parallel counters (GPCs), aiming to optimize the balance between logic utilization and delay. 
Kumm \textit{et al.} \cite{Datapath-2018TC-Kumm} further advanced heuristic method in \cite{Datapath-2008ASPDAC-Parandeh-Afshar, Datapath-2011TRETS-Parandeh-Afshar} and combined the heuristic method the with ILP.

Recently, machine learning methodologies have become promising solutions for circuit optimization and design space exploration, where various learning models are leveraged as surrogate models to evaluate designs during the search or optimization process \cite{DSE-2022TCAD-Geng,DSE-2019TCAD-MA,RL-2021DAC-Roy}.
An active learning-based prefix adder exploration framework is proposed in \cite{DSE-2019TCAD-MA}, which uses the Gaussian process regression model to predict the delay and area based on the feature extracted from the prefix tree structure. 
Geng \textit{et al.} \cite{DSE-2022TCAD-Geng} further facilitated automatic feature learning for prefix adder structures and deployed a sequential optimization framework that employs the graph neural process as a surrogate model, which enables a more efficient and effective adder structure exploration. 
However, the exploration still highly relies on a regression model as a proxy to the real PPA, whose modeling accuracy significantly affects the final results.
Contrary to existing approaches, reinforcement learning (RL) integrates actual PPA evaluations directly into its optimization loop, demonstrating its feasibility by efficiently navigating complex design spaces.
Recent advancements have seen RL tackle a variety of challenges within electronic design automation (EDA), as evidenced by applications across different domains such as prefix circuit design optimization, analog circuit design optimization and gate sizing \cite{RL-2021DAC-Roy,Sizing-2022DAC-Wang,Sizing-2022ICCAD-Siddharth}. 
Given the complexity of multiplier design and the vastness of its design space, the feasibility of reinforcement learning in multiplier design optimization is underscored. 
RL addresses this gap by leveraging real synthesized metrics as rewards, allowing optimization of designs that perform better than analytical models after synthesis.
In addition, RL algorithms can also be enhanced with parallelism by implementing different environment instances.
By utilizing multiple threads, the stability and efficiency of deep reinforcement learning algorithms are enhanced.

The design space of the multiplier is huge to explore. To address this, we have proposed an RL-based framework that is tailored for the optimization of multipliers and merged MACs~\cite{DSE-2023DAC-Zuo}. 
However, obtaining a suitable representation of the multiplier structure is non-trivial due to its inherent complexity.
To address this, we employ matrix and tensor representations for the compressor tree in a multiplier, enabling seamless integration of neural networks as the agent network in RL.
These representations effectively capture the structural characteristics of the multiplier.
The agent can learn to make effective decisions by optimizing the trade-off between key performance metrics such as power, performance, and area using a Pareto-driven approach.
Furthermore, to exploit the huge design space more efficiently, the proposed framework also features design space pruning and parallel RL agent training for more efficient optimization. 
To validate the effectiveness of the proposed framework, we applied it to design and optimize multipliers with different bit widths. The experimental results show that our approach outperforms various baseline methods, including legacy designs, evolutionary algorithms, and integer linear programming, in terms of area and delay.
Moreover, to validate the effectiveness of the optimized multipliers and MACs, a computation module, e.g., a process element (PE) array, is implemented with the multipliers and MACs generated by the RL agent, and the PPA gets improved accordingly. 
In summary, the contributions are as follows:
\begin{itemize}
    \item We propose a multiplier optimization framework based on reinforcement learning, marking the first instance of applying reinforcement learning for this purpose to our knowledge.
    \item 
    We present a matrix and a tensor representation for multipliers, which enables the seamless integration of deep neural networks as the agent network. A Pareto-driven reward is employed to accommodate the trade-off between the area and delay so that the agent can learn to achieve Pareto-optimal designs. 
    \item To improve search efficiency within this framework, we further enhance the framework with a parallel training methodology to enable faster and more stable training. 
    \item We also broaden the scope of the RL-based multiplier design framework to include fused MAC to validate the applicability.
    \item Experimental results demonstrate that the multipliers and MACs produced by RL agents dominate all baseline designs in terms of both area and delay. Furthermore, applying the optimized multipliers and MACs to the implementation of a larger computation module also results in PPA improvement, which validates the effectiveness of the optimized designs.
\end{itemize}

\section{Preliminary}
\label{sec:prelim}
\subsection{Multiplier Architecture}
\begin{figure}[!tb]
    \centering
    \includegraphics[width=.568\linewidth]{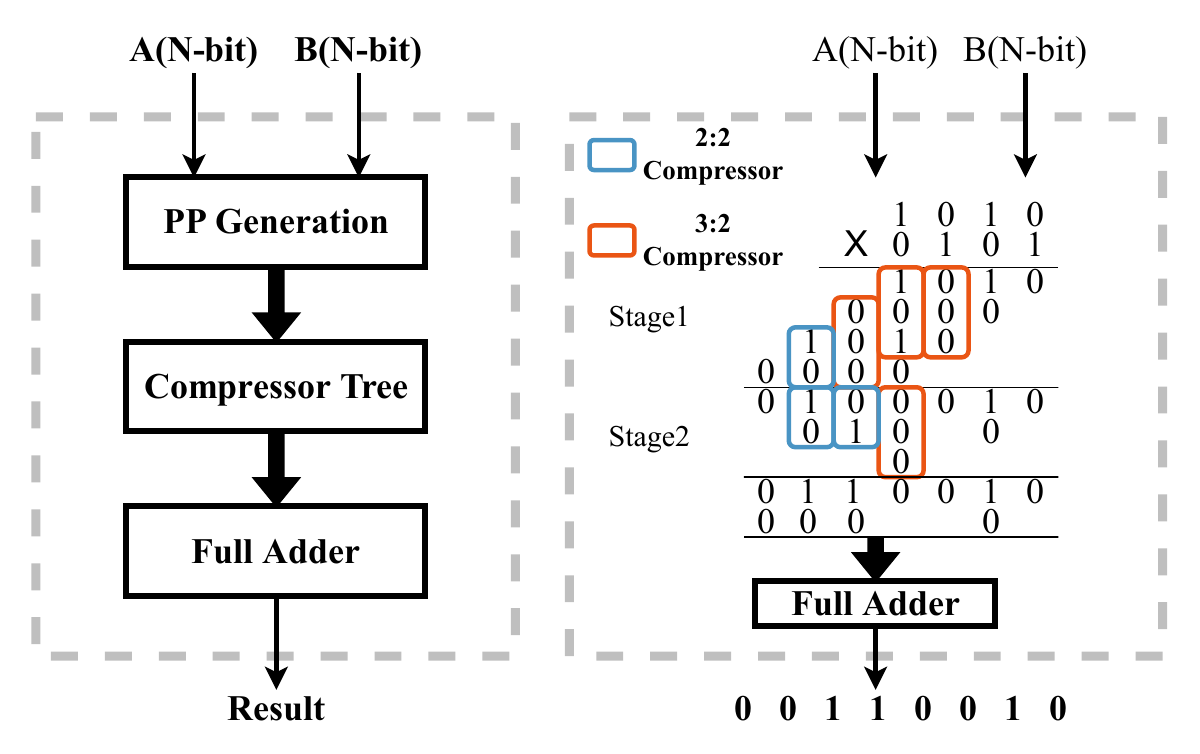}
    \caption{Multiplier architecture}
    \label{fig:multilpier}
\end{figure}

The multiplier typically comprises three primary components: a partial product generator (PPG), a compressor tree (CT), and a carry propagation adder, as shown in \Cref{fig:multilpier}. 
PPG generates partial products (PPs) from the multiplicand and multiplier, while the CT compresses these PPs into two parallel rows.
Subsequently, an adder is utilized to aggregate these two rows of PPs, culminating in the final product.
A typical partial product generator generally employs $N^2$ AND gates for an $N$-bit multiplier. 
A CT has multiple compression stages to compress the PPs into two rows. 
Predominantly, there are 3:2 compressors and 2:2 compressors implemented through a full adder and a half adder, respectively. 
A 3:2 (resp. 2:2) compressor applied at column $j$ of stage $i$ receives 3 (resp. 2) partial products as input from column $j$ of stage $i$, passing the \textit{sum} output to column $j$ of stage $i+1$, and the \textit{carry out} to column $j+1$ of stage $i+1$.   
Consequently, a 3:2 compressor decreases the partial products of column $j$ by two, while a 2:2 compressor reduces them by one, each incrementing the partial products in column $j+1$ by one.

\subsection{Q-Learning}
RL encompasses a collection of optimization problems referred to as \textit{state} $s$, with a corresponding set of \textit{actions} $A$. 
An agent transitions from one state $s$ to another state $s'$ by executing an action $a\in A$, consequently receiving a \textit{reward} $r(s,a)$ as an evaluation from the RL environment.
The model governing action selection is known as the \textit{policy} $\pi$. The primary objective of the RL agent is to devise a policy that optimizes the cumulative reward.

Q-learning is an RL algorithm that learns the scores of each action $a$ for a given state $s$, and the score is called Q-value, represented by $Q(s, a)$. 
According to Bellman equation~\cite{RL-1966Science-Bellman}, the Q-value is calculated as follows:
\begin{equation}
    Q\left( {s,a} \right) = r\left( {s,a} \right) + \gamma \mathop {\max }\limits_{a'} Q\left( {s',a'} \right),
\end{equation}
where $s'$ indicates the next state, and $\gamma$ is the discount factor. 
Therefore, the Q-value is updated by:
\begin{equation}
    Q\left( {s,a} \right) = Q\left( {s,a} \right) + \alpha \left[ {r\left( {s,a} \right) + \gamma \mathop {\max }\limits_{a'} Q\left( {s',a'} \right) - Q\left( {s,a} \right)} \right],
\end{equation}
where $\alpha$ is the learning rate. 
\label{subsec:Q}
In this paper, we utilize the deep Q-learning approach, leveraging a deep neural network to approximate the Q-value.
Here, the state $s$ represents the architecture of the multiplier, detailed in \Cref{subsec:ct-representation}.
An action $a$ alters the current multiplier architecture into a new one, effectively progressing to the next state.
The reward $r$ is quantified by the enhancements in the multiplier's area and delay metrics.

\subsection{Advantage Actor-Critic}
\label{subsec:a2c-prelim}
The A2C algorithm~\cite{RL-2016PMLR-A3C} addresses the challenges of high variance and unstable learning due to strong correlations between consecutive states by splitting the traditional RL model into two components: an actor that enacts policies and a critic that evaluates these actions. 
A policy network $\pi (a|s;\theta )$ and a value network $v(s;w)$ are employed, where $\theta$ and $w$ are the parameters of two neural networks, respectively. 
The synchronous multi-thread coordination method in A2C ensures uniform learning and parameter updates. This synchronization removes the need for different agents in A3C~\cite{RL-2016PMLR-A3C}, as the single agent with different environment instances suffices, while therefore avoiding updates based on outdated copies, significantly stabilizing the training process and offering sufficient parallelism and effectiveness ~\cite{winder2020reinforcement, jang2020study}.
The A2C algorithm employs bootstrapped advantage estimates generated by the critic instead of mere state-value approximations to enhance gradient estimation accuracy and learning efficiency~\cite{sewak2019deep}. 
With state $s$ and action $a$, The advantage is defined as:
\begin{equation}
    \hat{A}(s, a) = Q_{\pi}(s, a) - V_{\pi}(s),
\label{eq:origin_adv}
\end{equation}
where $Q_{\pi}(s, a)$ indicates the action-value function that estimates the expected reward that can be obtained by taking action $a$ and then following strategy $\pi$ at state $s$, while $V_{\pi}(s)$ indicates the state-value function that estimates the expected reward that can be obtained in state $s$ if the strategy $\pi$ is followed from that state instead of taking a specific action~\cite{winder2020reinforcement, RL-2016PMLR-Mnih}.

For a known transition $(s_t, a_t, r_t, s_{t+1})$, to facilitate calculations of advantage function, we use a value network $v(s_t;w)$ to approximate the state-value function $V_{\pi}(s_t)$, and estimate $Q_{\pi}(s_t, a_t)$ through Monte Carlo methods based on the Bellman equation. 
Consequently, we can approximate \eqref{eq:origin_adv} as:
\begin{equation}
    \hat{A}(s_t, a_t) \approx r_t + \gamma \cdot v(s_{t+1};w)-v(s_t;w),
    \label{eq:approx_adv}
\end{equation}
where $\gamma$ is the discount factor.

\section{Proposed Method}
\label{sec:algo}
\subsection{Overview}

\begin{figure}[!tb]
    \centering
    \includegraphics[width=0.75\linewidth]{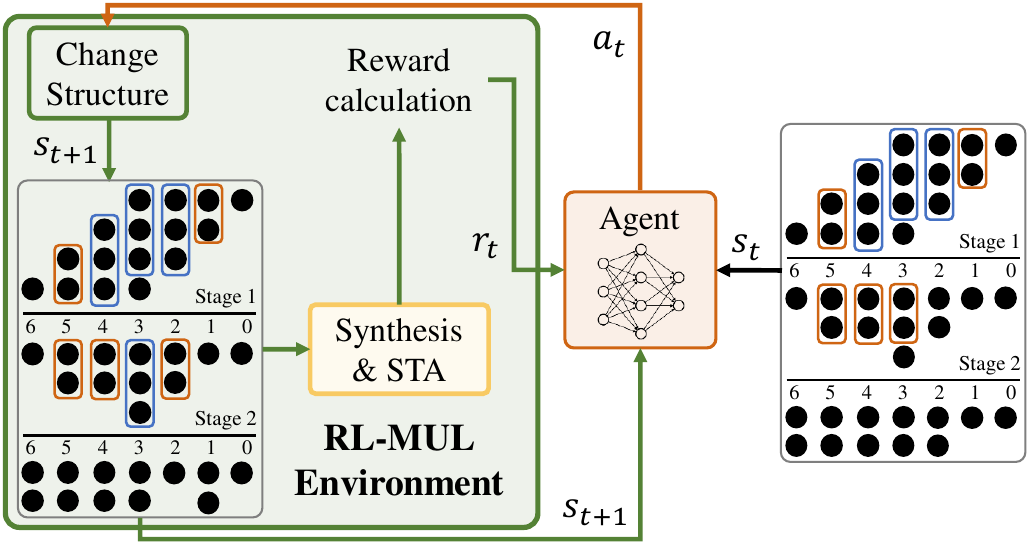}
    \caption{RL-MUL framework.}
    \label{fig:rl_flow}
\end{figure}

As illustrated in the left of \Cref{fig:rl_flow}, our original RL-MUL framework leverages a reinforcement learning approach for multiplier design optimization. 
An RL agent engages in iterative interactions with its environment from an initial state $s_0$.
At any given state $s_t$, the RL agent, guided by a policy $\pi$ derived from the policy network, selects an action $a_t$ from a set of legal actions.
This action modifies the current multiplier configuration, leading to a new state $s_{t+1}$. Subsequently, a reward $r_t$ is computed using EDA tools, facilitating the neural network model's update based on the received feedback.

\subsection{Multiplier Representation}
\label{subsec:ct-representation}

\begin{figure}[!tb]
    \centering
    \includegraphics[width=.4\linewidth]{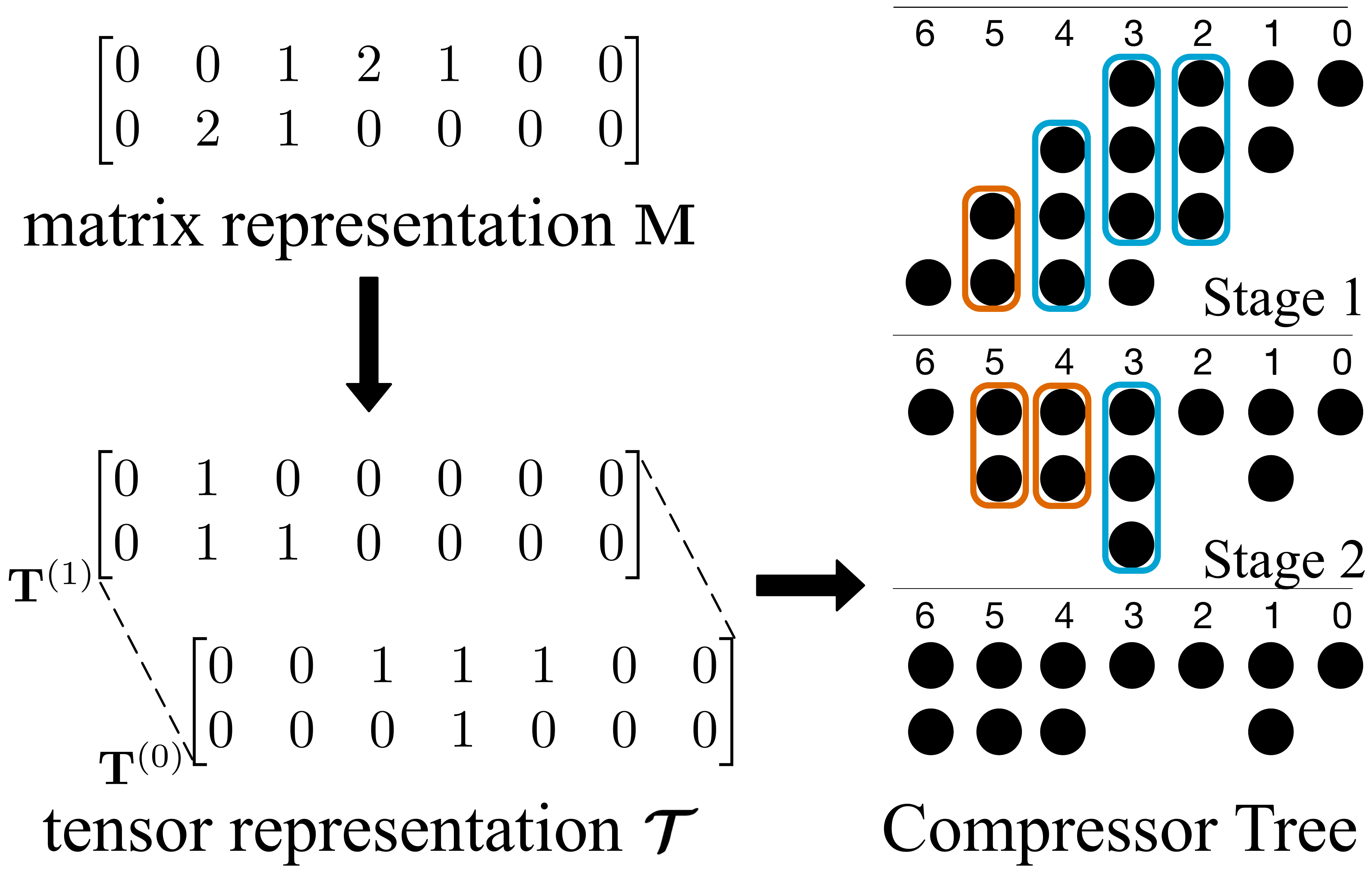}
    \caption{Structure representation.}
    \label{fig:ct-representation}
\end{figure}

The RL state space, denoted as $\mathcal{S}$, consists of all possible configurations of $N$-bit multipliers. 
We recognize the count of various compressors in each column as a critical attribute influencing the synthesized performance metrics of the multipliers. 
Consequently, we characterize the multiplier architecture using the aggregate counts of 3:2 and 2:2 compressors across columns, encapsulated by a matrix $\vec{M}\in \mathbb{R}^{2N\times 2}$. 
In this matrix, the first and second rows quantify the total 3:2 and 2:2 compressors in each column, respectively. 
An illustration of a 4-bit multiplier structure alongside its matrix representation $\vec{M}$ is provided in \Cref{fig:ct-representation}. 
To derive a complete multiplier structure from $\vec{M}$, the compressors are allocated to specific stages. 
However, the mapping from $\vec{M}$ to the structures is not unique since different assignments of compressors in multiple stages may have the same overall number in each column.
To achieve a distinctive representation, we advance to a tensor representation that offers more informative insights, as illustrated in \Cref{fig:ct-representation}.

We represent the tensor as $\mathbfcal{T} \in \mathbb{R}^{K \times 2N \times ST}$, with $K$ indicating the total kinds of compressors used and $ST$ the stage count. 
Specifically, we utilize 3:2 and 2:2 compressors, thus $K=2$. 
This framework is designed for potential extension to accommodate more compressor variants. 
The tensors $\vec{T}^{(0)}=\ten{T}_{0,:,:} \in \mathbb{R}^{2N\times ST}$ and $\vec{T}^{(1)}=\ten{T}_{1,:,:}\in \mathbb{R}^{2N\times ST}$ respectively map the placement of 3:2 and 2:2 compressors. 
The elements $t^{(0)}_{ij}$ and $t^{(1)}_{ij}$ denote the quantity of 3:2 and 2:2 compressors at the $j$-th column of the $i$-th stage. 
Given a matrix $\vec{M}$ that contains the information of the overall number of compressors in each column, we can construct the tensor representation $\ten{T}$ correspondingly based on an assignment scheme of the compressors in different stages. 

For the assignment process, we employ a deterministic method that assigns compressors from the least to the most significant bit columns, prioritizing 3:2 compressors and then utilizing 2:2 compressors where applicable. 
This method progresses through stages until all compressors are allocated, as detailed in \Cref{alg:assignment}. 
This approach guarantees a unique tensor representation for each multiplier structure, facilitating precise and unambiguous characterizations of the multiplier architecture.

\begin{algorithm}[!tb]
\caption{Compressor Assignment}\label{alg:assignment}
\begin{algorithmic}[1]
\Require
$\vec{M}$: Matrix representation 
\Ensure 
$\mathbfcal{T}$: Tensor representation.
\For{$j \gets 1\ to\ 2N$}
    \State {$i \gets 0$}
    \While{column $j$ exists not assigned comp.}
        \State{Assign 3:2 comp. to stage $i$ column $j$ first} \label{alg:assignment:32-1}
        \State{Update $t^{(0)}_{ij}$in $\vec{T}^{(0)}$} \label{alg:assignment:32-2}
        \If{Remaining PPs $\geq2$} \label{alg:assignment:22-1}
            \State{Assign 2:2 comp. to stage $i$ column $j$} \label{alg:assignment:22-2}
            \State{Update $t^{(1)}_{ij}$in $\vec{T}^{(1)}$} \label{alg:assignment:22-3}
        \EndIf

        \State {$i \gets i+1$}
    \EndWhile
\EndFor
\State {$\ten{T}_{0,:,:} \gets \vec{T}^{(0)}$}
\State {$\ten{T}_{1,:,:} \gets \vec{T}^{(1)}$}

\end{algorithmic}
\end{algorithm}

\begin{figure}[!tb]
    \centering
    \includegraphics[width=\linewidth]{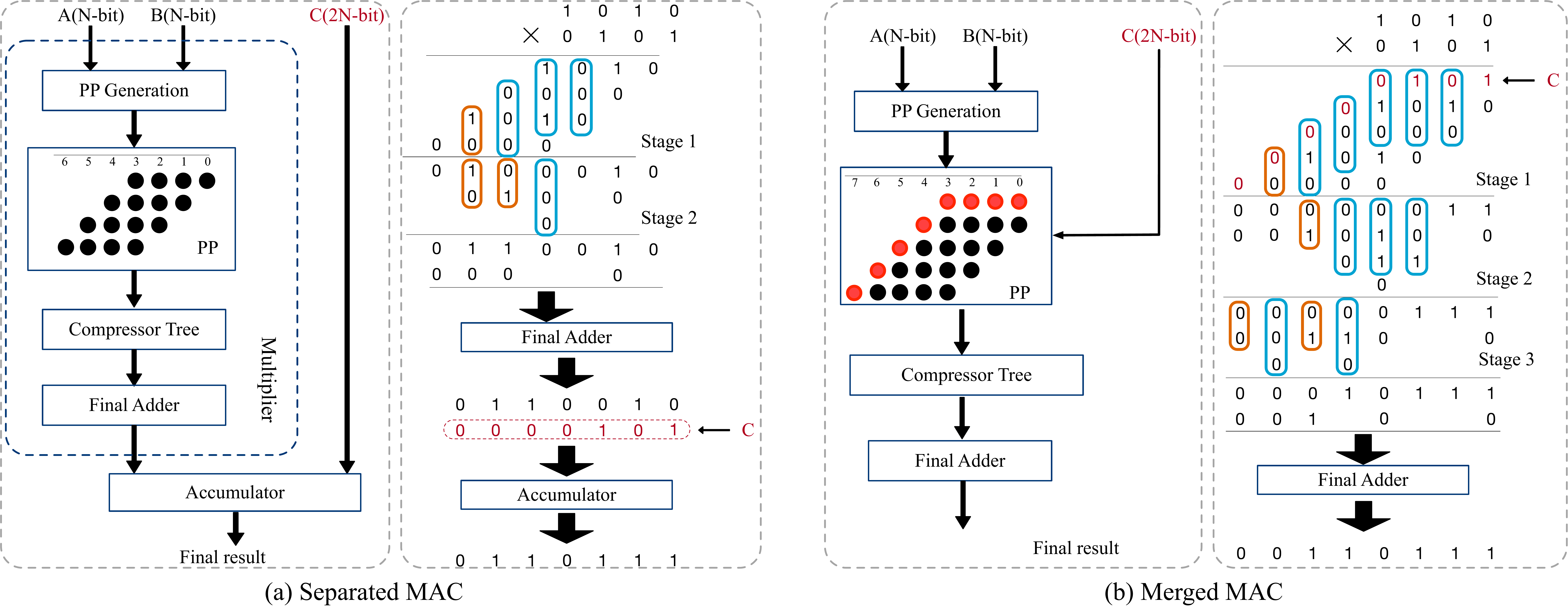}
    \caption{MAC architectures}
    \label{fig:mac}
\end{figure}

\subsection{Multiplier Modification}
\label{subsec:action}

In the RL agent's context, an action $a$ signifies the agent's choice to alter the existing structure of the multiplier. 
The agent can choose from four distinct actions for each column: adding or removing a 2:2 compressor, and replacing a 3:2 or a 2:2 compressor with another type. 
We denote $res_j$ to present the PP number after compression of column $j$, which should only be 1 or 2.
Actions leading to $res_j$ values of 0 or 3, such as adding or removing a 3:2 compressor, are excluded, thereby defining the action space as $|\mathcal{A}|=2N\times4=8N$. 
It is important to note that not every action is feasible to yield a legal multiplier structure instantly. 
For instance, if there is no 2:2 compressor in column $1$ as depicted in \Cref{fig:ct-representation}, attempting to remove a 2:2 compressor from this column is considered invalid.
Similarly, any action $a$ on column $j$ that results in the partial product (PP) numbers post-compression being either 0 or 3 would be considered invalid. 
Take another example from \Cref{fig:ct-representation}. 
Removing a 2:2 compressor from column $4$ would lead to $res_4$ equating to three, thereby invalidating the action.

For a compressor tree with $2N$ columns, the output of a deep Q-network is a vector that indicates the predicted Q-values:
\begin{equation}
\hspace{-.18in}
    Q(s_t)=[\textcolor{myorange}{q_{11}},\textcolor{myorange}{q_{12}},\textcolor{myorange}{q_{13}},\textcolor{myorange}{q_{14}},\cdots,\textcolor{myyellow}{q_{2N,1}},\textcolor{myyellow}{q_{2N,2}}, \textcolor{myyellow}{q_{2N,3}}\textcolor{black}{,}\textcolor{myyellow}{q_{2N,4}}],
    \label{equation:2N_q}
\end{equation}
where each group of $q_{j1}, q_{j2}, q_{j3}, q_{j4}$ indicates the Q-value of the four actions $a_{j1}, a_{j2}, a_{j3}, a_{j4}$ in column $j$.
\label{subsec:mask}
To ensure only legal actions can be selected, a mask $\vec{m}$ is utilized as the selector to enable valid actions and forbid invalid actions. 
\begin{equation}
\hspace{-.2in}
    \vec{m} = [\textcolor{myorange}{m_{10}},\textcolor{myorange}{m_{11}},\textcolor{myorange}{m_{12}},\textcolor{myorange}{m_{13}},\cdots,\textcolor{myyellow}{m_{2N,0}},\textcolor{myyellow}{m_{2N,1}}, \textcolor{myyellow}{m_{2N,2}}\textcolor{black}{,}\textcolor{myyellow}{m_{2N,3}}],
\end{equation}
where each entry is a binary value.
If an action $a_{ij}$ is valid, the corresponding entry in $m_{ij}$ is 1. Otherwise, it is 0.
In the proposed RL framework, the final masked Q-value vector is the element-wise multiplication of the mask vector and Q-value vector: 
\begin{equation}
    Q'(s_t) = Q(s_t)\odot\vec{m}.
\label{eq:masked_q}
\end{equation}
Now the decision is given by 
\begin{equation}
{a_t} = \mathop {\arg \max }\limits_a Q'\left( {{s_t},a} \right).
\label{eq:decision_q}
\end{equation}

Note that only non-zero entries are considered. 
The action applied to column $j$ changes the number of 3:2 or 2:2 compressors of the current column $j$,
which may cause the number of compressed PPs of subsequent column $j+1$ to become 0 or 3.
We use the legalization strategy shown in \Cref{alg:legalization} to refine the multiplier structure to ensure the PPs are compressed to 2 rows. 
This strategy sequentially refines from column $j+1$ to the MSB, addressing under-compression by adding or replacing compressors, and managing over-compression by removing compressors. 
Similar to the assignment procedure, the legalization process is also deterministic. 
Under state $s_t$, we can get a new state $s_{t+1}$ after performing action $a_t$ to modify the structure along with the legalization.

\subsection{Pareto-driven Reward}
\label{subsec:reward}

In our framework, we define the reward, $r_t$, as the improvement in circuit metrics, such as area, delay, and power, achieved by executing an action $a_t$ at state $s_t$.
Considering the nature of the trade-off between power, performance, and area (PPA), a superior multiplier design is always expected to achieve Pareto-optimal in terms of these dimensions. 
To encourage the RL agent to learn to generate Pareto-optimal designs, we introduce a Pareto-driven reward mechanism. 
This mechanism leverages a synthesis flow under multiple design constraints, enabling the reward to cover a variety of design scenarios: those driven primarily by area, delay, or power, as well as scenarios seeking a trade-off optimization of these three key metrics.
The overall cost is calculated as a weighted sum of area, delay, and power, allowing for flexible adjustment of their relative importance in different design scenarios:
\begin{equation}
cost = w_{a}\sum_{i=1}^{n} area_i + w_{d}\sum_{i=1}^{n} delay_i + w_{p}\sum_{i=1}^{n} power_i,
\label{eq:cost}
\end{equation}
where  $area_i$, $delay_i$, and $power_i$ are the synthesized metrics under the $i$-th constraint. 
Since the area, delay, and power values have substantially different ranges, we normalize the area, delay, and power metrics to a consistent scale using Wallace tree implementations.
$w_{a}$, $w_{d}$ and $w_{p}$ are the weights to trade off PPA.
We define our reward $r$ as the difference between $s_t$ and $s_{t+1}$:
\begin{equation}
r_t = cost_t - cost_{t+1}
\label{eq:reward}
\end{equation}

\begin{algorithm}[!tb]
\caption{Legalization}\label{alg:legalization}
\begin{algorithmic}[1]
\Require Multiplier structure to be legalized; $C$: action column
\Ensure Legalized multiplier structure
    \For{$j \gets (C+1)\ to\ 2N$} 
        \State {$res_j \gets$ Get residual PPs after compression}
        \If{$res_j = 1$ or $res_j = 2$}
            \State \Return \Comment{legalization done}
        \ElsIf{$res_j == 3$}  \label{alg:legalization:3-1}
            \If{exists 2:2 comp. in column $j$} \label{alg:legalization:3-2}
                \State {Replace a 2:2 compressor}\label{alg:legalization:3-3}
            \Else \label{alg:legalization:3-4}
                \State {Add a 3:2 compressor} \label{alg:legalization:3-5}
            \EndIf
        \ElsIf{$res_j == 0$}  \label{alg:legalization:0-1}
            \If{exists 2:2 compressor in column $j$} \label{alg:legalization:0-2}
                \State {Delete a 2:2 compressor} \label{alg:legalization:0-3}
            \Else \label{alg:legalization:0-4}
                \State {Delete a 3:2 compressor} \label{alg:legalization:0-5}
            \EndIf \label{alg:legalization:0-6}
        \EndIf
    \EndFor
\end{algorithmic}
\end{algorithm}

\subsection{Training Algorithm}
\label{subsec:training}

We adopt ResNet-18~\cite{CV-2016CVPR-He} as the backbone of Q-Network
with the parameters denoted by $\theta$. 
The state undergoes encoding into a tensor representation $\ten{T}$, as detailed in \Cref{subsec:ct-representation}, before being processed by the Q-network. 
The RL training methodology 
is outlined in \Cref{alg:dqn}. 
Initially, action selections $a$ are randomized during the warm-up phase (\Cref{alg:dqn:random}), transitioning to policy-based selections in subsequent steps (\Cref{alg:dqn:policy}). 

Each iteration $t$ leads to the transformation of the multiplier's architecture from $s_t$ to $s_{t+1}$, culminating in a reward $r_t$ derived from synthesis and timing analysis. 
This process generates a new transition $(s_t, a_t, r_t, s_{t+1})$, which is 
recorded.
Then, the network parameter $\theta$ is updated by gradient descent, and $w$ is also updated in actor-critic methods(\Cref{alg:dqn:update}).
The target Q-value for each state-action pair within the batch is determined as follows:
\begin{equation}
    y = r' + \gamma \max\limits_{a'}Q'(s',a';\theta),
    \label{equation:target-q-value}
\end{equation}
where $\gamma$ is the discount factor. Based on the expected Q-value $y$, a gradient of $\theta$ can be obtained by:
\begin{equation}
    \Delta \theta =  \nabla_\theta(y-Q'(s,a;\theta))^2.
    \label{equation:q_theta-update}
\end{equation}
Then, the network parameter $\theta$ is updated by gradient descent(\Cref{alg:dqn:update}). By incorporating masked actions in backpropagation, the Q-network learns to assign lower Q-values to seldom-used, invalid actions, minimizing their selection in future iterations.

\begin{algorithm}[!tb]
\caption{RL-MUL flow
}\label{alg:dqn}
\begin{algorithmic}[1]
\Require 
$\vec{M_0}$: initial multiplier structure; 
$\gamma$: discount factor; 
$\alpha$: learning rate;
$T$: total training steps;
$T_B$: warm-up steps
\Ensure
$\theta$: Q-network parameters
\State{Replay buffer $B\gets \{\}$}
\State{Encode $s_0$ into $\ten{T}$ based on $\vec{M_0}$ \Comment{\Cref{alg:assignment}}}
\State{$t\gets0$}
\For{$t\gets0$ to $T$}
    \If{$t < T_B$}
        \State{$a_t\gets$ randomly choose from legal actions} \label{alg:dqn:random}
    \Else
        \State{Get $a_t$ by \Cref{eq:decision_q}}
        \label{alg:dqn:policy}
    \EndIf
    \State{Perform $a_t$ to $s_t$ and get $s_{t+1}$} \label{alg:dqn:tran}
    \State{Run EDA tools on $s_{t+1}$ and get $r_t$ \Comment{\Cref{eq:reward}}} \label{alg:dqn:eda}
    \State{Push $(s_t,a_t,r_t,s_{t+1}$) to $B$} \label{alg:dqn:push}
    \State{Sample a batch of transitions from $B$} \label{alg:dqn:sample}
    \State{Update $\theta$ by gradient descent \Comment{\Cref{equation:q_theta-update}}} \label{alg:dqn:update}
\EndFor

\end{algorithmic}
\end{algorithm}

\section{RL-MUL 2.0}
\label{sec:efficient}
Optimizing hardware configurations requires an efficient search within a vast design space, particularly in multiplier design, where an increase in bit width exponentially expands the design space.
Therefore, dealing with this enlarged space effectively becomes crucial, especially for larger designs like MACs, where the DQN algorithm may struggle to achieve optimal results.

In this work, we extend the proposed RL framework to MAC designs, enhancing its application in deep learning acceleration. To tackle the greater challenges of more complex designs, we use parallel algorithms to improve efficiency from two perspectives. 
Firstly, parallel optimization is always a promising solution in this scenario. 
Their inherent parallelism not only reliably boosts search efficiency but also fosters a thorough exploration of possible designs, enhancing the likelihood of uncovering optimal or nearly optimal solutions.
Secondly, search space pruning condenses the design space by discarding less promising designs. 
This approach emphasizes exploring viable design areas, thus refining the search process and minimizing computational demands. 
Eliminating inferior designs early on ensures a more targeted and efficient discovery of superior configurations.
In addition, integrating metrics with high correlation can achieve a similar purpose, not only significantly simplifying the optimization process but also enhancing the focus on configurations that genuinely contribute to performance improvements.

\subsection{Extend to Merged Multiply-Accumulator Architecture}
\label{subsec:mac-extend}
An integral component of many digital signal processing systems and neural network architectures is the multiply accumulator (MAC), which can be a decisive factor in determining the overall performance of many computing-intensive systems.
Incorporating compressor trees within MACs offers a pathway to enhance their efficiency. 
Rather than treating multiplication and accumulation as sequential operations, this approach seamlessly integrates them. 
By merging the accumulation (addition) directly into the partial product stages of multiplication and conducting partial product compression, we can capitalize on parallelism, thus potentially speeding up the entire MAC operation.
We can see that the proposed RL framework can seamlessly support the optimization of fused MAC design.

By integrating the addition into the partial product generation phase, the representation within the RL framework is tweaked to consider the intricacies of the MAC operation. 
The aim is to train the RL agent to explore and design optimal MAC structures, utilizing compressor trees for efficient parallel addition.
Therefore, the representations in \Cref{subsec:ct-representation} can be easily extended to MACs by providing ``merged" partial products, and the training procedure will be identical. 
In \Cref{sec:exp}, we will demonstrate the effectiveness and superiority of the proposed RL framework for fused MAC design. 

\begin{figure}[!tb]
    \centering
    \includegraphics[width=0.616\linewidth]{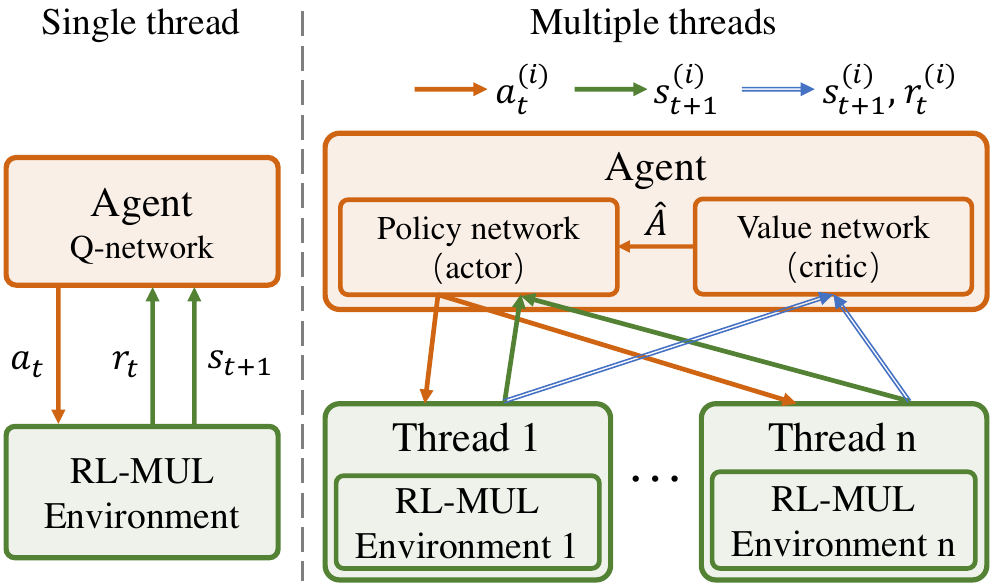} 
    \caption{Comparison of RL algorithm between single-thread and multi-thread implementations.}
    \label{fig:rlmul1_vs_rlmul2}
\end{figure}

\subsection{Multiple Agents Training}
\label{subsec:parallel}
Multiple agents running in parallel are more likely to explore different parts of the environment, promoting more efficient and stable policy training.
Therefore, as shown in~\Cref{fig:rlmul1_vs_rlmul2}, compared to the single-thread RL algorithm implemented in the proposed RL framework, we further enhance the proposed framework by training multiple agents in parallel, where each agent is handled by a thread. 
Following the training stability analysis in~\cite{RL-2016PMLR-A3C}, each agent in our framework is designed with the A2C scheme, where the policy and value networks share the convolution layers of ResNet-18.
Specifically, we employ a shared global network parameter across threads, each interacting with its local environment independently, followed by an average of all threads' gradient updates to adjust the global parameter. 
$n$ parallel threads synchronously process corresponding transitions $(s_t^{(i)}, a_t^{(i)}, r_t^{(i)}, s_{t+1}^{(i)})$. 
The right side of \Cref{fig:rl_flow} illustrates the synchronous parallel structure with A2C. 
At each step, given $n$ is the number of threads, the agent selects $a_{t}$ including $n$ actions for threads, the $i$-th thread's RL-MUL environment instance receives its corresponding action $a_{t}^{(i)} $ and transitions to the new state $s_{t+1}^{(i)} $, which is returned with the reward $r_{t}^{(i)} $ to the agent. 
Thus, in A2C, each element of the transition tuples $(s_t, a_t, r_t, s_{t+1})$ is an $n$-element vector.
Then \Cref{equation:2N_q} is transformed into:
\begin{equation}
    \pi(\cdot |s_t)=[\textcolor{myorange}{\pi_{11}},\textcolor{myorange}{\pi_{12}},\textcolor{myorange}{\pi_{13}},\textcolor{myorange}{\pi_{14}},\cdots,\textcolor{myyellow}{\pi_{2N,1}},\textcolor{myyellow}{\pi_{2N,2}}, \textcolor{myyellow}{\pi_{2N,3}}\textcolor{black}{,}\textcolor{myyellow}{\pi_{2N,4}}],
\end{equation}
where each group of $\pi_{j1}, \pi_{j2}, \pi_{j3}, \pi_{j4}$ indicates the probability of the four actions $a_{j1}, a_{j2}, a_{j3}, a_{j4}$ in column $j$.
The mask is configured as in \Cref{subsec:mask}, and the final masked probability distribution vector is: 
\begin{equation}
    \pi'(\cdot |s_t) = \pi(\cdot |s_t)\odot\vec{m}.
\end{equation}
Now, the decision is given by 
\begin{equation}
a_t \sim \pi'(\cdot |s_t).
\label{eq:decision_a2c}
\end{equation}

\Cref{alg:a2c} outlines the parallel training and optimization flow in RL-MUL 2.0~\cite{sewak2019deep, winder2020reinforcement}.
Firstly, $n$ threads (\Cref{alg:a2c:thread}) are initiated. Then, at each step, a multiplier structure alteration $a_t$ is sampled from \Cref{eq:decision_a2c}, which incorporates masks to prevent the selection of actions that lead to invalid multiplier structures (\Cref{alg:a2c:policy}). Following this, the chosen action is executed to obtain a new structure and its corresponding reward (\Cref{alg:a2c:tran}).
In terms of the model updating, this algorithm employs an $n$-step return approach for faster learning, updating the policy network parameter $\theta $ and the value network parameter $w$ only when the total training steps constitute an integer multiple of the update interval (\Cref{alg:a2c:Tup}).
A policy gradient of $\theta$ used to update the policy network can be obtained by:
\begin{equation}
    \Delta\theta = \nabla_\theta \log \pi(a_t|s_t;\theta) \cdot \hat{A}(s_t, a_t),
    \label{equation:theta-update}
\end{equation}
where $\pi(a_t|s_t;\theta)$ is the policy network definited by parameter $\theta$ at time $t$, and $\hat{A}$ is the advantage function defined in ~\Cref{eq:approx_adv}~\cite{RL-2016PMLR-Mnih}.
Then, the policy network parameter $\theta$ is updated by gradient ascent (\Cref{alg:a2c:update1}).
In addition, the Temporal-Difference (TD) learning~\cite{RL-1994PhDThesis-Multi-Step-Q-Learning} aspect of the A2C algorithm guides the value network $v(s_{t};w)$ to converge to the TD target $y_t$, defined as:
\begin{equation}
    y_t= r_t + \gamma \cdot v(s_{t+1};w),
\end{equation}
where $\gamma$ represents the discount factor. The TD target combines the real reward $r_t$ after taking the action $a_t$ with the predicted value of the next state $s_{t+1}$, serving as a crucial element in computing the TD error. This error is expressed as:
\begin{equation}
\delta _{t}  =v(s_{t};w)-y_{t} ,
\end{equation}
measuring the discrepancy between the estimated value of the state before taking the action $a_t$ and the TD target. In other words, the TD error reflects the accuracy of the value function prediction.
Based on the TD error, a gradient of $w$ used to update the value network can be obtained by:
\begin{equation}
\Delta w = -\nabla_w \frac{(\delta _t)^2}{2} =-\delta _t\cdot \nabla_w v(s_{t};w) .
\label{equation:w-update}
\end{equation}
Then, the value network parameter $w$ is updated by gradient descent (\Cref{alg:a2c:update2}). 


\begin{algorithm}[!tb]
\caption{RL-MUL 2.0 flow}\label{alg:a2c}
\begin{algorithmic}[1]
\Require
$n$: number of threads;
$T$: total training steps;
$t_{up}$: update interval
\Ensure
$\theta$: policy network parameters;
$w$: value network parameters
\State{Initialize $n$ parallel threads} \label{alg:a2c:thread} 
\For{$t\gets0$ to $T$} 
\State{Sample a multiplier structure alteration \(a_{t}^{(i)} \ \text{by \Cref{eq:decision_a2c}, } \forall i \in \{1, 2, \ldots, n\}\)}\label{alg:a2c:policy}
\State{Perform $a_t^{(i)}$ to get structure $s_{t+1}^{(i)}$ and $r_t^{(i)}$, $\forall i \in \{1, 2, \ldots, n\}$}\label{alg:a2c:tran}
\If{$t \mid t_{up}$}\label{alg:a2c:Tup}
\State{Update $\theta$ by gradient ascent \Comment{\Cref{equation:theta-update}}} 
\label{alg:a2c:update1}
\State{Update $w$ by gradient descent \Comment{\Cref{equation:w-update}}} \label{alg:a2c:update2}
\EndIf
\EndFor
\end{algorithmic}
\end{algorithm}

\begin{figure}[!tb]
    \centering
    \includegraphics[width=0.916\linewidth]{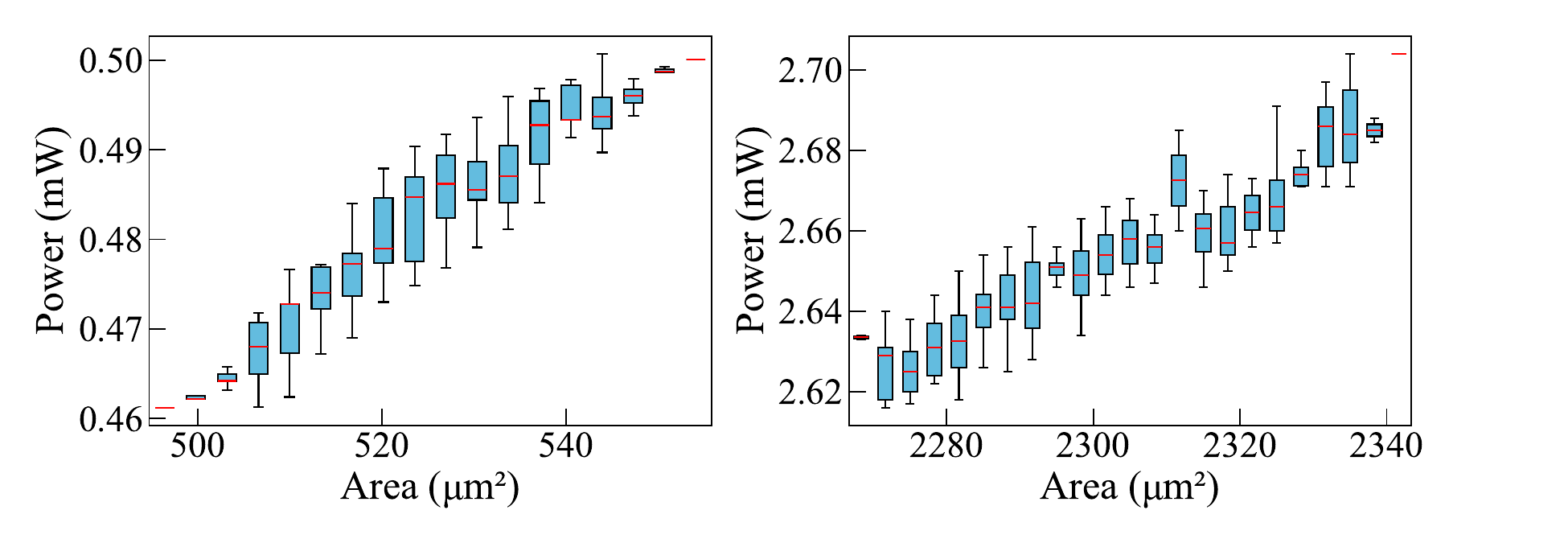} 
    \vspace{-.2in}
    \caption{Correlation between area and power. The upper one is an 8-bit AND-based multiplier, and the lower one is a 16-bit AND-based multiplier.}
    \label{fig:areavspower}
\end{figure}

\subsection{Objective Space Reduction}
\label{subsec:pruning}
The goal of multiplier design space exploration is to find designs that are superior in terms of multiple objectives. 
In \Cref{eq:reward}, a weighted reward is designed such that the agent can acquire a good trade-off among different objectives, while the selection of weights for each objective can impact the final solutions substantially. 
The more objectives we have, the more effort is required for tuning the weights.
Notably, we investigated a correlation between the area and the power of a multiplier.
Based on the architectures we have searched for, it is observed that the power and area are highly correlated. 
A correlation between these two factors represented by box plots is illustrated in \Cref{fig:areavspower}, the upper graph depicts the relationship for 8-bit AND-based multipliers,
while the lower plot shows the same for 16-bit AND-based multipliers. 
The bottom and top boundaries of the box represent the first and third quartiles, respectively, indicating the inter-quartile range (IQR). 
The median is denoted by the band within the box. 
The upper whisker represents the maximum value of the data, and the lower whisker represents the minimum value of the data.
It can be observed from the trend in \Cref{fig:areavspower} that there exists a strong positive correlation between the area and the power, which suggests that the area is a reliable indicator of the power.
Consequently, our methodology gives precedence to area and delay as key optimization metrics, which allows \Cref{eq:cost} to be further reduced to:
\begin{equation}
cost = w_{a}\sum_{i=1}^{n} area_i + w_{d}\sum_{i=1}^{n} delay_i
\label{eq:new_cost}
\end{equation}


\subsection{Search Space Pruning}
\label{sec:search-space-pruning}
\begin{figure}[!tb]
    \centering
    \includegraphics[width=0.86\linewidth]{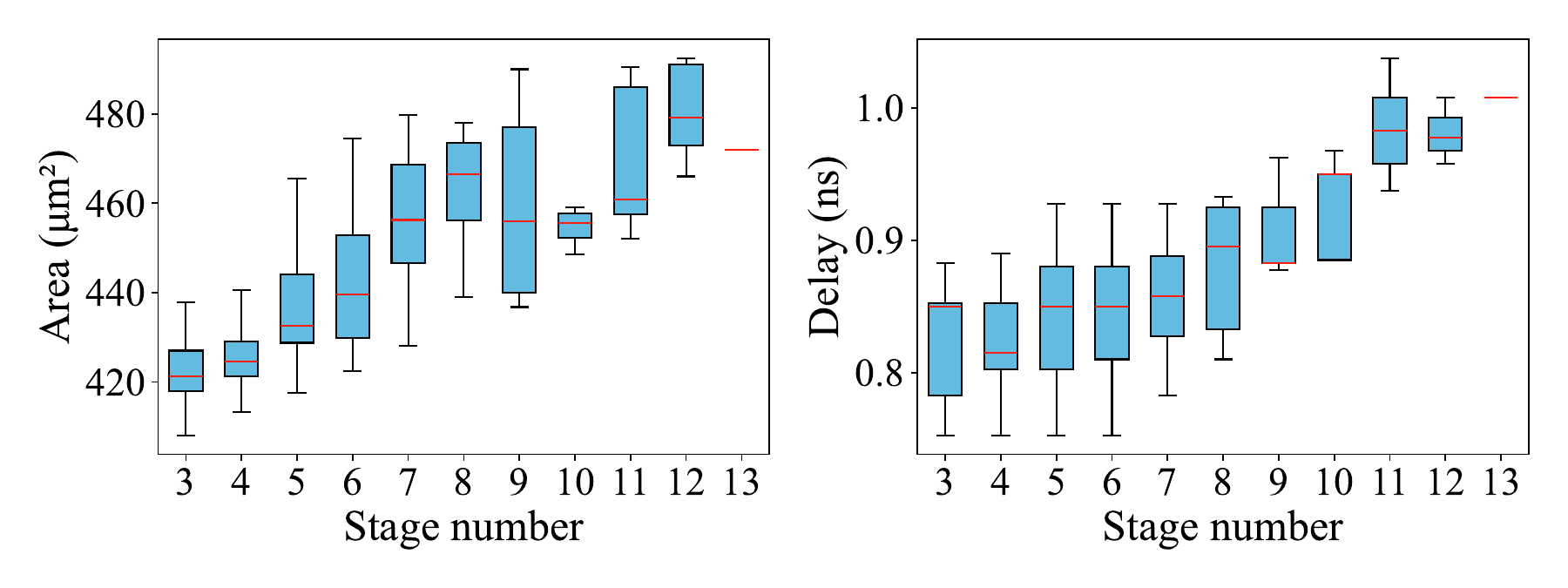}
    \vspace{-.2in}
    \caption{Correlation between stage number and metrics of 8-bit AND-based multipliers.}
    \label{fig:stage_relation}
\end{figure}

Furthermore, another analysis indicates the number of stages of a compressor tree as a significant factor affecting the area and delay of multipliers, as shown in \Cref{fig:stage_relation}. This analysis takes 8-bit AND-based multiplier structures as an example. 
Notably, there is a positive relationship between stage number and the parameters of area and delay. This suggests that an increase in stage number is associated with a corresponding rise in these metrics.
To mitigate this, our framework integrates a strategy to constrain actions that will lead to excessive stage increases, which facilitates a more efficient search and optimization toward desired multiplier structures. 


\section{Experimental Results}
\label{sec:exp}
\subsection{Setup}
\label{subsec:exp-setup}

The proposed framework is implemented on a Linux system powered by a 2.8 GHz AMD EPYC CPU and an NVIDIA RTX 3090 GPU.
We use EasyMAC \cite{Datapath-2022ASPDAC-Jiaxi} for RTL generation and have extended its capabilities by incorporating Modified Booth Encoding (MBE)-based partial product generation, as well as enhancing support for the RTL generation of merged MAC units.
These designs are synthesized using the OpenROAD flow~\cite{OpenROAD-2019GOMACTECH-Ajayi} with the NanGate $45nm$ Open Cell Library~\cite{nangate45}.  
OpenSTA\cite{OpenSTA} is utilized to perform timing analysis. 
To ensure the functional correctness of the generated multipliers, we first convert RTL into AIGER format using Yosys \cite{Yosys}, and use the \texttt{cec} command in ABC\cite{Berkeley-ABC} to perform logic equivalence verification with a golden implementation of multiplier. 

Given the prevalent use of 8-bit and 16-bit multipliers, the RL-MUL 2.0 framework is assessed using both 8-bit and 16-bit multipliers, incorporating AND-based PPG and MBE-based PPG.
We compare our approach against established baselines, including the legacy Wallace tree\cite{Datapath-1964TC-Wallace}, an ILP-based method GOMIL \cite{Datapath-2021DATE-Xiao}, and the simulated annealing (SA) technique.
Four delay constraints are configured in \Cref{eq:cost}. 
The weights $w_{a}$ and $w_{d}$ range from 0 to 1, resulting in different optimization preferences towards area or delay. 
In native RL-MUL implementation, we set $\gamma$ to 0.8, learning rate to 0.0002, $\epsilon$ to decay from 0.95 to 0.05, and employ RMSProp optimizer \cite{RMSProp} for the training.
In the RL-MUL 2.0 implementation, we employ four synchronization threads and a five-step return.
We train the original RL-MUL and RL-MUL 2.0 10,000$s$ and run SA for the same amount of time. During this period, RL-MUL performs approximately 450 iterations, while RL-MUL 2.0 performs about 650 iterations, exploring 4 design points in each iteration. Each iteration of RL-MUL takes about 22 seconds, and each iteration of RL-MUL 2.0 takes about 15 seconds.
For the ILP approach, solving the 8-bit cases takes approximately 1 $min$, whereas the 16-bit cases require around 16 $h$.
Synthesizing under varying design constraints produces different netlists for the same RTL design. 
We synthesize all the obtained multipliers and MACs across target delays from 0.05 $ns$ to 1.2 $ns$.
Furthermore, to enhance the demonstration of RL-MUL 2.0's effectiveness and the performance of the resulting designs, we incorporate these multipliers and MACs from all evaluated methods into large macro designs.
Processing Element (PE) arrays, commonly utilized in DNN accelerators, consist of numerous MAC units, making them ideal for further evaluating the impact of different multipliers and MACs on area and timing efficiency.
By integrating different multipliers and MACs into PE arrays, specifically following a systolic array architecture, we investigate the potential for improvements in both area and timing within these structures. 

\begin{figure}
    \centering


\begin{tikzpicture}
\begin{groupplot}[group style={group size= 4 by 1,  horizontal sep=1.4cm, group name=myplot},height=3.2cm,width=3.8cm]
\nextgroupplot[minor tick num=0,
ymax=0.9,
xlabel={Area ($\upmu$m$^2$)},
ylabel={Delay (ns)},
y label style={at={(-0.2,0.5)}},
ylabel near ticks,
legend style={
    draw=none,
	at={(0.5,1.15)},
	nodes={scale=0.75, transform shape},
	anchor=north,
	legend columns=-1,
}
]   
    \addplot[blue, style={mark=*, mark size=0.7pt, line width=0.8pt, draw=mydarkblue, draw opacity=0.7, mark options={fill opacity=0.7}}]  table [x=area, y=delay, col sep=comma] {pgfplot/data/processed/Processed-mult_wallace_8.csv};
    \addplot[blue, style={mark=*, mark size=0.7pt, line width=0.8pt, draw=myblue, draw opacity=0.7, mark options={fill opacity=0.7}}]  table [x=area, y=delay, col sep=comma] {pgfplot/data/processed/Processed-mult_ilp_8.csv};
    \addplot[blue, style={mark=*, mark size=0.7pt, line width=0.8pt, draw=myyellow, draw opacity=0.7, mark options={fill opacity=0.7}}]  table [x=area, y=delay, col sep=comma] {pgfplot/data/final_multi_constraint_sa_8bit.csv};
    \addplot[blue, style={mark=*, mark size=0.7pt, line width=0.8pt, draw=myorange, draw opacity=0.7, mark options={fill opacity=0.7}}]  table [x=area, y=delay, col sep=comma] {pgfplot/data/final_multi_constraint_dqn_8bit.csv};
    \addplot[blue, style={mark=*, mark size=0.7pt, line width=0.8pt, draw=mygreen, draw opacity=0.7, mark options={fill opacity=0.7}}]  table [x=area, y=delay, col sep=comma] {pgfplot/data/final_multi_constraint_a2c_8bit.csv};
\coordinate (left) at (rel axis cs:0,1);

\nextgroupplot[minor tick num=0,
xlabel={Area ($\upmu$m$^2$)},
ylabel={Delay (ns)},
y label style={at={(-0.2,0.5)}},
xtick={600,700},
ymin=0.88,ymax=1.13,
ytick={0.9,1.0,1.1},
ylabel near ticks,
legend style={
    draw=none,
	at={(1.0,1.35)},
	nodes={scale=0.75, transform shape},
	anchor=north,
	legend columns=-1,
}
]   
    \addplot[blue, style={mark=*, mark size=0.7pt, line width=0.8pt, draw=mydarkblue, draw opacity=0.8, mark options={fill opacity=0.8}}]  table [x=area, y=delay, col sep=comma] {pgfplot/data/processed/Processed-mult_booth_wallace_8.csv};\addlegendentry{Wallace~\cite{Datapath-1964TC-Wallace}};
    \addplot[blue, style={mark=*, mark size=0.7pt, line width=0.8pt, draw=myblue, draw opacity=0.8, mark options={fill opacity=0.8}}]  table [x=area, y=delay, col sep=comma] {pgfplot/data/processed/Processed-mult_booth_ilp_8.csv};\addlegendentry{GOMIL~\cite{Datapath-2021DATE-Xiao}};
    \addplot[blue, style={mark=*, mark size=0.7pt, line width=0.8pt, draw=myyellow, draw opacity=0.8, mark options={fill opacity=0.8}}]  table [x=area, y=delay, col sep=comma] {pgfplot/data/final_multi_constraint_sa_8bit_booth.csv};\addlegendentry{SA};
    \addplot[blue, style={mark=*, mark size=0.7pt, line width=0.8pt, draw=myorange, draw opacity=0.8, mark options={fill opacity=0.8}}]  table [x=area, y=delay, col sep=comma] {pgfplot/data/final_multi_constraint_dqn_8bit_booth.csv};\addlegendentry{RL-MUL~\cite{DSE-2023DAC-Zuo}};
    \addplot[blue, style={mark=*, mark size=0.7pt, line width=0.8pt, draw=mygreen, draw opacity=0.8, mark options={fill opacity=0.8}}]  table [x=area, y=delay, col sep=comma] {pgfplot/data/final_multi_constraint_a2c_8bit_booth.csv};\addlegendentry{RL-MUL 2.0};
\coordinate (mid) at (rel axis cs:0.5,1);

\nextgroupplot[minor tick num=0,
xlabel={Area ($\upmu$m$^2$)},
ylabel={Delay (ns)},
xtick={1800,2200},
xmax=2500,
y label style={at={(-0.2,0.5)}},
ylabel near ticks,
legend style={
    draw=none,
	at={(0.9,1.2)},
	nodes={scale=0.75, transform shape},
	anchor=north,
	legend columns=-1,
}
]
    \addplot[blue, style={mark=*, mark size=0.7pt, line width=0.8pt, draw=mydarkblue, draw opacity=0.8, mark options={fill opacity=0.8}}]  table [x=area, y=delay, col sep=comma] {pgfplot/data/processed/Processed-mult_wallace_16.csv};
    \addplot[blue, style={mark=*, mark size=0.7pt, line width=0.8pt, draw=myblue, draw opacity=0.8, mark options={fill opacity=0.8}}]  table [x=area, y=delay, col sep=comma] {pgfplot/data/processed/Processed-mult_ilp_16.csv};
    \addplot[blue, style={mark=*, mark size=0.7pt, line width=0.8pt, draw=myyellow, draw opacity=0.8, mark options={fill opacity=0.8}}]  table [x=area, y=delay, col sep=comma] {pgfplot/data/final_multi_constraint_sa_16bit.csv};
    \addplot[blue, style={mark=*, mark size=0.7pt, line width=0.8pt, draw=myorange, draw opacity=0.8, mark options={fill opacity=0.8}}]  table [x=area, y=delay, col sep=comma] {pgfplot/data/final_multi_constraint_dqn_16bit.csv};
    \addplot[blue, style={mark=*, mark size=0.7pt, line width=0.8pt, draw=mygreen, draw opacity=0.8, mark options={fill opacity=0.8}}]  table [x=area, y=delay, col sep=comma] {pgfplot/data/final_multi_constraint_a2c_16bit.csv};
\coordinate (right) at (rel axis cs:1,1);

\nextgroupplot[minor tick num=0,
xlabel={Area ($\upmu$m$^2$)},
ylabel={Delay (ns)},
xtick={2000,2400},
y label style={at={(-0.2,0.5)}},
ylabel near ticks,
legend style={
    draw=none,
	at={(0.5,1.15)},
	nodes={scale=0.75, transform shape},
	anchor=north,
	legend columns=-1,
}
]
    \addplot[blue, style={mark=*, mark size=0.7pt, line width=0.8pt, draw=mydarkblue, draw opacity=0.8, mark options={fill opacity=0.8}}]  table [x=area, y=delay, col sep=comma] {pgfplot/data/processed/Processed-mult_booth_wallace_16.csv};
    \addplot[blue, style={mark=*, mark size=0.7pt, line width=0.8pt, draw=myblue, draw opacity=0.8, mark options={fill opacity=0.8}}]  table [x=area, y=delay, col sep=comma] {pgfplot/data/processed/Processed-mult_booth_ilp_16.csv};
    \addplot[blue, style={mark=*, mark size=0.7pt, line width=0.8pt, draw=myyellow, draw opacity=0.8, mark options={fill opacity=0.8}}]  table [x=area, y=delay, col sep=comma] {pgfplot/data/final_multi_constraint_sa_16bit_booth.csv};
    \addplot[blue, style={mark=*, mark size=0.7pt, line width=0.8pt, draw=myorange, draw opacity=0.8, mark options={fill opacity=0.8}}]  table [x=area, y=delay, col sep=comma] {pgfplot/data/final_multi_constraint_dqn_16bit_booth.csv};
    \addplot[blue, style={mark=*, mark size=0.7pt, line width=0.8pt, draw=mygreen, draw opacity=0.8, mark options={fill opacity=0.8}}]  table [x=area, y=delay, col sep=comma] {pgfplot/data/final_multi_constraint_a2c_16bit_booth.csv};
\coordinate (right) at (rel axis cs:1.5,1);

\end{groupplot}
\path (left)--(right) coordinate[midway] (group center);
\path (myplot c2r1.north west|-current bounding box.north)--
coordinate(legendpos)
(myplot c2r1.north west|-current bounding box.north);
\end{tikzpicture}
    \vspace{-.2in}
    \caption{ Pareto-frontiers of the synthesis results on multipliers. From left to right: 8-bit AND-based; 8-bit MBE-based; 16-bit AND-based; 16-bit MBE-based. Note that the Pareto-frontiers of RL-MUL 2.0 and original RL-MUL overlap in the 8-bit MBE-based result.}
    \label{fig:results-mult}
\end{figure}
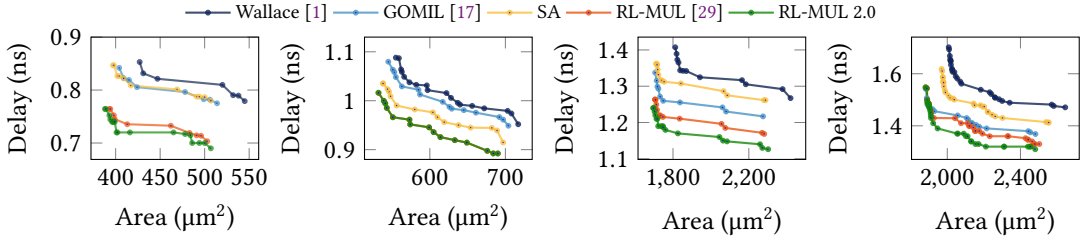

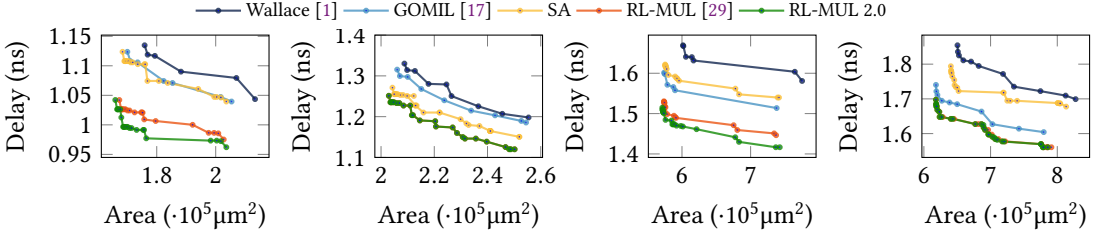
\begin{figure}
    \centering
    \usepgfplotslibrary{groupplots}


\begin{tikzpicture}
\begin{groupplot}[group style={group size= 4 by 1,  horizontal sep=1.4cm, group name=myplot},height=3.2cm,width=3.8cm]
\nextgroupplot[minor tick num=0,
xlabel={Area ($\cdot 10^5\upmu$m$^2$)},
ylabel={Delay (ns)},
y label style={at={(-0.2,0.5)}},
ylabel near ticks,
legend style={
    draw=none,
	at={(0.5,1.15)},
	nodes={scale=0.75, transform shape},
	anchor=north,
	legend columns=-1,
}
]
    \addplot[blue, style={mark=*, mark size=0.7pt, line width=0.8pt, draw=mydarkblue, draw opacity=0.8, mark options={fill opacity=0.8}}]  table [x=area, y=delay, col sep=comma] {pgfplot/data/pe/pe_mult_8_wallace.csv};
    \addplot[blue, style={mark=*, mark size=0.7pt, line width=0.8pt, draw=myblue, draw opacity=0.8, mark options={fill opacity=0.8}}]  table [x=area, y=delay, col sep=comma] {pgfplot/data/pe/pe_mult_8_ilp.csv};
    \addplot[blue, style={mark=*, mark size=0.7pt, line width=0.8pt, draw=myyellow, draw opacity=0.8, mark options={fill opacity=0.8}}]  table [x=area, y=delay, col sep=comma] {pgfplot/data/pe/pe_mult_8bit_and_sa_0103.csv};
    \addplot[blue, style={mark=*, mark size=0.7pt, line width=0.8pt, draw=myorange, draw opacity=0.8, mark options={fill opacity=0.8}}]  table [x=area, y=delay, col sep=comma] {pgfplot/data/pe/pe_mult_8bit_and_dqn_0103.csv};
    \addplot[blue, style={mark=*, mark size=0.7pt, line width=0.8pt, draw=mygreen, draw opacity=0.8, mark options={fill opacity=0.8}}]  table [x=area, y=delay, col sep=comma] {pgfplot/data/pe/pe_mult_8bit_and_a2c_0103_merged.csv};
\coordinate (left) at (rel axis cs:0,1);

\nextgroupplot[minor tick num=0,
ymin=1.1, ymax=1.4,
xlabel={Area ($\cdot 10^5\upmu$m$^2$)},
ylabel={Delay (ns)},
y label style={at={(-0.2,0.5)}},
ylabel near ticks,
legend style={
    draw=none,
	at={(1.0,1.35)},
	nodes={scale=0.75, transform shape},
	anchor=north,
	legend columns=-1,
}
]
    \addplot[blue, style={mark=*, mark size=0.7pt, line width=0.8pt, draw=mydarkblue, draw opacity=0.8, mark options={fill opacity=0.8}}]  table [x=area, y=delay, col sep=comma] {pgfplot/data/pe/pe_mult_booth_wallace_8.csv}; \addlegendentry{Wallace~\cite{Datapath-1964TC-Wallace}};
    \addplot[blue, style={mark=*, mark size=0.7pt, line width=0.8pt, draw=myblue, draw opacity=0.8, mark options={fill opacity=0.8}}]  table [x=area, y=delay, col sep=comma] {pgfplot/data/pe/pe_mult_booth_ilp_8.csv};\addlegendentry{GOMIL~\cite{Datapath-2021DATE-Xiao}};
    \addplot[blue, style={mark=*, mark size=0.7pt, line width=0.8pt, draw=myyellow, draw opacity=0.8, mark options={fill opacity=0.8}}]  table [x=area, y=delay, col sep=comma] {pgfplot/data/pe/pe_mult_8bit_booth_sa_0108.csv};\addlegendentry{SA};
    \addplot[blue, style={mark=*, mark size=0.7pt, line width=0.8pt, draw=myorange, draw opacity=0.8, mark options={fill opacity=0.8}}]  table [x=area, y=delay, col sep=comma] {pgfplot/data/pe/pe_mult_8bit_booth_dqn_0103.csv};\addlegendentry{RL-MUL~\cite{DSE-2023DAC-Zuo}};
    \addplot[blue, style={mark=*, mark size=0.7pt, line width=0.8pt, draw=mygreen, draw opacity=0.8, mark options={fill opacity=0.8}}]  table [x=area, y=delay, col sep=comma] {pgfplot/data/pe/pe_mult_8bit_booth_a2c_0103.csv};\addlegendentry{RL-MUL 2.0};
\coordinate (mid) at (rel axis cs:0.5,1);

\nextgroupplot[minor tick num=0,
xlabel={Area ($\cdot 10^5\upmu$m$^2$)},
ylabel={Delay (ns)},
y label style={at={(-0.2,0.5)}},
ylabel near ticks,
legend style={
    draw=none,
	at={(1,1.35)},
	nodes={scale=0.75, transform shape},
	anchor=north,
	legend columns=-1,
}
]
    \addplot[blue, style={mark=*, mark size=0.7pt, line width=0.8pt, draw=mydarkblue, draw opacity=0.8, mark options={fill opacity=0.8}}]  table [x=area, y=delay, col sep=comma] {pgfplot/data/pe/pe_mult_16_wallace.csv};
    \addplot[blue, style={mark=*, mark size=0.7pt, line width=0.8pt, draw=myblue, draw opacity=0.8, mark options={fill opacity=0.8}}]  table [x=area, y=delay, col sep=comma] {pgfplot/data/pe/pe_mult_16_ilp.csv}; 
    \addplot[blue, style={mark=*, mark size=0.7pt, line width=0.8pt, draw=myyellow, draw opacity=0.8, mark options={fill opacity=0.8}}]  table [x=area, y=delay, col sep=comma] {pgfplot/data/pe/pe_mult_16bit_and_sa_0103.csv};
    \addplot[blue, style={mark=*, mark size=0.7pt, line width=0.8pt, draw=myorange, draw opacity=0.8, mark options={fill opacity=0.8}}]  table [x=area, y=delay, col sep=comma] {pgfplot/data/pe/pe_mult_16bit_and_dqn_0103.csv};
    \addplot[blue, style={mark=*, mark size=0.7pt, line width=0.8pt, draw=mygreen, draw opacity=0.8, mark options={fill opacity=0.8}}]  table [x=area, y=delay, col sep=comma] {pgfplot/data/pe/pe_mult_16bit_and_a2c_0103_merged.csv};
\coordinate (right) at (rel axis cs:1,1);

\nextgroupplot[minor tick num=0,
xlabel={Area ($\cdot 10^5\upmu$m$^2$)},
ylabel={Delay (ns)},
y label style={at={(-0.2,0.5)}},
ylabel near ticks,
legend style={
    draw=none,
	at={(0.5,1.15)},
	nodes={scale=0.75, transform shape},
	anchor=north,
	legend columns=-1,
}
]
    \addplot[blue, style={mark=*, mark size=0.7pt, line width=0.8pt, draw=mydarkblue, draw opacity=0.8, mark options={fill opacity=0.8}}]  table [x=area, y=delay, col sep=comma] {pgfplot/data/pe/pe_mult_booth_wallace_16.csv};
    \addplot[blue, style={mark=*, mark size=0.7pt, line width=0.8pt, draw=myblue, draw opacity=0.8, mark options={fill opacity=0.8}}]  table [x=area, y=delay, col sep=comma] {pgfplot/data/pe/pe_mult_booth_ilp_16.csv};
    \addplot[blue, style={mark=*, mark size=0.7pt, line width=0.8pt, draw=myyellow, draw opacity=0.8, mark options={fill opacity=0.8}}]  table [x=area, y=delay, col sep=comma] {pgfplot/data/pe/pe_mult_16bit_booth_sa_0103.csv};
    \addplot[blue, style={mark=*, mark size=0.7pt, line width=0.8pt, draw=myorange, draw opacity=0.8, mark options={fill opacity=0.8}}]  table [x=area, y=delay, col sep=comma] {pgfplot/data/pe/pe_mult_16bit_booth_dqn_0122_merged.csv};
    \addplot[blue, style={mark=*, mark size=0.7pt, line width=0.8pt, draw=mygreen, draw opacity=0.8, mark options={fill opacity=0.8}}]  table [x=area, y=delay, col sep=comma] {pgfplot/data/pe/pe_mult_16bit_booth_a2c_0122_merged.csv};
\coordinate (right) at (rel axis cs:1.5,1);
\end{groupplot}

\path (left)--(right) coordinate[midway] (group center);
\path (myplot c3r1.north west|-current bounding box.north)--
coordinate(legendpos)
(myplot c3r1.north west|-current bounding box.north);
\end{tikzpicture}
    \vspace{-.2in}
    \caption{Pareto-frontiers of the synthesis results on multiplier-implemented PE arrays. From left to right: 8-bit AND-based; 8-bit MBE-based; 16-bit AND-based; 16-bit MBE-based.}
    \label{fig:results-pe}
\end{figure}

\begin{figure}[!tb]
    \centering
    \subfloat[]{\pgfplotsset{
	width =0.8\linewidth,
	height=0.22\linewidth
}

\begin{tikzpicture}
\begin{axis}[
    ybar,
    xticklabels={8-bit-AND, 8-bit-MBE, 16-bit-AND, 16-bit-MBE},xtick={0.8,2.3,3.8,5.3},
    xtick align=inside,
    ylabel={Normalized HV},
    ylabel near ticks,
    bar width = 6pt,
    xmin=0,
    xmax=6.1,
    ymin=0,
    ymax=1.1,
    legend style={at={(0.5,1.2)},
    draw=none,anchor=south,legend columns=-1,font=\small}
    ]

\addplot +[ybar, fill=mydarkblue,   draw=black, area legend] table [x={x},  y={wallace}] {hv.dat};
\addplot +[ybar, fill=myblue,   draw=black, area legend] table [x={x},  y={ilp}] {hv.dat};
\addplot +[ybar, fill=myyellow,   draw=black, area legend] table [x={x},  y={sa}] {hv.dat};
\addplot +[ybar, fill=myorange,   draw=black, area legend] table [x={x},  y={dqn}] {hv.dat};
\addplot +[ybar, fill=mygreen,   draw=black, area legend] table [x={x},  y={a2c}] {hv.dat};
\legend{Wallace~\cite{Datapath-1964TC-Wallace}, GOMIL~\cite{Datapath-2021DATE-Xiao}, SA, RL-MUL~\cite{DSE-2023DAC-Zuo}, RL-MUL 2.0}

\end{axis}
\end{tikzpicture} \label{fig:hv_mul}} \hspace{.01in}
    \subfloat[]{\pgfplotsset{
	width =0.8\linewidth,
	height=0.22\linewidth
}

\begin{tikzpicture}
\begin{axis}[
    ybar,
    xticklabels={8-bit-AND, 8-bit-MBE, 16-bit-AND, 16-bit-MBE},xtick={0.8,2.3,3.8,5.3},
    xtick align=inside,
    ylabel={Normalized HV},
    ylabel near ticks,
    bar width = 6pt,
    xmin=0,
    xmax=6.1,
    ymin=0,
    ymax=1.1,
    legend style={at={(0.5,1.35)},
    draw=none,anchor=north,legend columns=-1}
    ]

\addplot +[ybar, fill=mydarkblue,   draw=black, area legend] table [x={x},  y={wallace}] {hv_pe_mul.dat};
\addplot +[ybar, fill=myblue,   draw=black, area legend] table [x={x},  y={ilp}] {hv_pe_mul.dat};
\addplot +[ybar, fill=myyellow,   draw=black, area legend] table [x={x},  y={sa}] {hv_pe_mul.dat};
\addplot +[ybar, fill=myorange,   draw=black, area legend] table [x={x},  y={dqn}] {hv_pe_mul.dat};
\addplot +[ybar, fill=mygreen,   draw=black, area legend] table [x={x},  y={a2c}] {hv_pe_mul.dat};
\end{axis}
\end{tikzpicture} \label{fig:hv_pe}}
    \vspace{-.1in}
    \caption{Pareto-frontiers hypervolume comparison of (a) multipliers and (b) multiplier-implemented PE arrays.}
\end{figure}
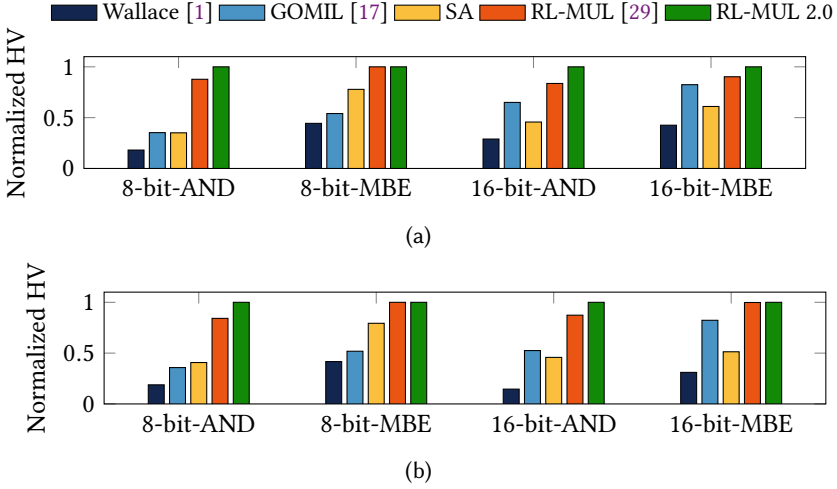

\subsection{Multiplier Performance Comparison}


The resulting area-delay curves for multipliers are illustrated in \Cref{fig:results-mult}, where the designs derived from the RL-MUL framework outperform all baselines.
Detailed statistics of minimum area, delay, and balanced area-delay metrics are presented in \Cref{table:mult-results} (the optimal results are marked in bold). 
In the trade-off scenario, optimal corresponds to the lowest weighted sum of PPA in \Cref{eq:reward}.
Through the RL-MUL 2.0 framework, we achieve up to 10.0\% area reduction under the minimum area constraint and a 12.5\% decrease in delay under the minimum delay constraint.
Additionally, the implementation in PE arrays, as shown in \Cref{fig:results-pe} and \Cref{table:pe-results}, indicates similar performance, with up to a 6.0\% area reduction and up to 11.5\% delay decrease.

The hypervolume~\cite{Zitzler-2007EMO-Zitzler} measures the volume enclosed by the Pareto frontier and a reference point in the objective space, which is a common metric to evaluate the quality of the Pareto frontiers. 
Hypervolume comparisons for multipliers presented in \Cref{fig:hv_mul} show that RL-MUL generates significantly larger hypervolume than GOMIL, with average increases of 85.9\%. 
RL-MUL 2.0 shows an improvement of 11.1\% compared to the original RL-MUL. 
Similarly, for PE arrays constructed with the multiplier, as shown in \Cref{fig:hv_pe}, the average improvement of RL-MUL 2.0 compared with GOMIL and original RL-MUL is 96.1\% and 8.4\% respectively.

\begin{table}[!tb]
\caption{Multiplier area, timing, and power comparison.}
\vspace{-.1in}
\centering
\resizebox{\linewidth}{!} {
\begin{tabular}{c|c|ccc|ccc|ccc|ccc}
    \hline
    \multirow{3}{*}{Preference} &\multirow{3}{*}{Method}& \multicolumn{6}{c|}{8-bit} & \multicolumn{6}{c}{16-bit} \\ \cline{3-14}
                      &      & \multicolumn{3}{c|}{AND} & \multicolumn{3}{c|}{MBE} & \multicolumn{3}{c|}{AND} & \multicolumn{3}{c}{MBE} \\ 
                        &    & Area ($\upmu$m$^2$) & Delay (ns) & Power (mW) & Area ($\upmu$m$^2$) & Delay (ns) & Power (mW) & Area ($\upmu$m$^2$) & Delay (ns) & Power (mW) & Area ($\upmu$m$^2$) & Delay (ns) & Power (mW)  \\ \hline
    \multirow{4}{*}{Area} & Wallace\cite{Datapath-1964TC-Wallace}&  427   &   0.8530  & 0.3513 &  555  &   1.0880  & 0.4975 &  1812   &   1.4073 & 1.962 &  2008    &  1.7016 & 2.187\\
    &GOMIL \cite{Datapath-2021DATE-Xiao}    &  404   &   0.8420  & 0.3352 & 545  &  1.0797 & 0.4833 &  1706   &   1.3375 & 1.855 &   1882   & 1.5432 & 2.013\\ 
    &SA   &  397   &   0.8468  & 0.3317 &  538 &  1.0353  &  0.4845&   1712    & 1.3619  &1.860  & 1969 & 1.6184 & 2.133\\  
    &\textbf{RL-MUL}   &  393   &   0.7643  &0.3261  & \textbf{532}  &  \textbf{1.0162}  &  \textbf{0.4752}&   1705    & 1.2633  &1.855  & 1882  & 1.5478 & 2.016\\  
    &\textbf{RL-MUL 2.0}   &  \textbf{388}   &   \textbf{0.7643}  &  \textbf{0.3237}& \textbf{532}  &   \textbf{1.0162}  &  \textbf{0.4752}&   \textbf{1696}    & \textbf{1.2481}  & \textbf{1.845} & \textbf{1881} & \textbf{1.5478} & \textbf{2.008}\\  \hline
    
    \multirow{4}{*}{Timing} & Wallace\cite{Datapath-1964TC-Wallace} &  545   &   0.7791  & 0.4977 &  720  &   0.9601  & 0.7054 &  2420   &   1.2672  & 2.822 &  2645 & 1.4709 & 3.032\\
    &GOMIL \cite{Datapath-2021DATE-Xiao} & 514    &  0.7750  & 0.4726 &  706    &  0.9571 & 0.6836 &  2281   &   1.2169  & 2.629 &   2482   & 1.3684 & 2.791\\ 
    &SA   &  507   &   0.7800  & 0.4656 &  697 &  0.9147  &  0.6886&    2280   & 1.2616  &2.619  & 2551  & 1.4125 & 2.893\\  
    &\textbf{RL-MUL}   &  503   &   0.7033  &0.4650  &  \textbf{690} &  \textbf{0.8922}  &  \textbf{0.6736}&   2281    &  1.1684  &2.638  &  2475 &  1.3318 & 2.780\\  
    &\textbf{RL-MUL 2.0}   &  \textbf{507}   &  \textbf{0.6931}  & \textbf{0.4670} &  \textbf{690} &    \textbf{0.8922}  &  \textbf{0.6736}&   \textbf{2302}    &   \textbf{1.1263}  & \textbf{2.658} &  \textbf{2481} & \textbf{1.3085} & \textbf{2.791}\\  \hline

    \multirow{4}{*}{Trade-off} & Wallace \cite{Datapath-1964TC-Wallace}&  458   &   0.8328 & 0.3820 & 637 &   1.0018  & 0.5900 &  2184   &   1.3054  & 2.562 &  2300 & 1.4954 & 2.537\\
    &GOMIL\cite{Datapath-2021DATE-Xiao}  &435 &  0.8086  & 0.3634 &  629 &  0.9837 & 0.5824 &  2061   &   1.2416  & 2.382 &   2106   & 1.4298 & 2.328\\ 
    &SA   &   402  &   0.8265  & 0.3366 & 556  &   0.9901  &  0.5163&   1738    & 1.3161  &1.907  &  2016 & 1.5071 & 2.232\\  
    &\textbf{RL-MUL}   &  399   &   0.7451  &0.3345  & \textbf{551}  &   \textbf{0.9662}  &  \textbf{0.5081}&   1731    &  1.2192  &1.903  &  1927 & 1.4339 & 2.140\\  
    &\textbf{RL-MUL 2.0}   &   \textbf{401}  &   \textbf{0.7252}  & \textbf{0.3360} &  \textbf{551} &   \textbf{0.9662}  &  \textbf{0.5081}&    \textbf{1722}   &  \textbf{1.1875} & \textbf{1.887} &  \textbf{1947} & \textbf{1.3923} & \textbf{2.148}\\  
    \hline
\end{tabular}
}
\label{table:mult-results}
\end{table}

\begin{table}[!tb]
\caption{PE array (multiplier) area, timing, and power comparison.}
\vspace{-.1in}
\centering
\resizebox{\linewidth}{!} {
\begin{tabular}{c|c|ccc|ccc|ccc|ccc}
    \hline
    \multirow{3}{*}{Preference} &\multirow{3}{*}{Method}& \multicolumn{6}{c|}{8-bit} & \multicolumn{6}{c}{16-bit} \\ \cline{3-14}
                      &      & \multicolumn{3}{c|}{AND} & \multicolumn{3}{c|}{MBE} & \multicolumn{3}{c|}{AND} & \multicolumn{3}{c}{MBE} \\ 
                        &    & Area ($\upmu$m$^2$) & Delay (ns) & Power (mW) & Area ($\upmu$m$^2$) & Delay (ns) & Power (mW) & Area ($\upmu$m$^2$) & Delay (ns) & Power (mW) & Area ($\upmu$m$^2$) & Delay (ns) & Power (mW) \\ \hline
    \multirow{4}{*}{Area} & Wallace\cite{Datapath-1964TC-Wallace}&175892  &   1.1347   & 145.14 &   208782   &   1.3302  & 178.67 &   601492  &   1.6693   & 495.67 &  650385    & 1.8543  & 548.16 \\ 
    &GOMIL\cite{Datapath-2021DATE-Xiao}        &  170036   &   1.1237  & 141.11 &   206058   &   1.3154  & 176.36 &   574117  &   1.6017  & 470.92 &  618107    & 1.7403  & 522.13 \\ 
    &SA   &  168401   &   1.1237  & 140.08 & 204288  &  1.2711  &  175.17&   575479     &  1.6216  &472.59  & 640443   &  1.794  & 536.34 \\  
    &\textbf{RL-MUL}   &  167312    &  1.0421   & 138.79 &  \textbf{202926}  &   \textbf{1.2512}  &  \textbf{173.18} &    573709   & 1.5305   &471.27  & 618107   &  1.6976  & 520.65 \\  
    &\textbf{RL-MUL 2.0}   &  \textbf{165950}   &  \textbf{1.0421}   & \textbf{137.65} & \textbf{202926}  &   \textbf{1.2512}  & \textbf{173.18} &   \textbf{571394}   & \textbf{1.5148}   & \textbf{469.23} &  \textbf{617971} &  \textbf{1.6976}  & \textbf{520.57} \\  \hline
    
    \multirow{4}{*}{Timing} & Wallace\cite{Datapath-1964TC-Wallace} &  213345   &   1.0436   & 175.97 &  258016    &  1.1988  & 220.73 &   775001  &   1.5809  & 639.48 &  827503  &  1.6992  & 692.68 \\ 
    &GOMIL \cite{Datapath-2021DATE-Xiao}       &  205378   &   1.0395   & 169.73 &  254475    &   1.1856  & 217.20 &  739591   &   1.5137  & 610.51 &  785896    & 1.6085  & 659.21 \\ 
    &SA   &  203607    &  1.0395   & 168.23 &  251955 &   1.1505  & 214.85 &  739318    &  1.5398  &608.93  & 813339   & 1.6777  & 681.73 \\  
    &\textbf{RL-MUL}   &  202722    &  0.9752   & 167.61 &  \textbf{250185}  &  \textbf{1.1200}  &  \textbf{213.30} &  737139    & 1.4464  &606.31  & \textbf{778678}   & \textbf{1.5607}  & \textbf{652.30} \\  
    &\textbf{RL-MUL 2.0}   & \textbf{203607}    &  \textbf{0.9621}   & \textbf{168.52} &  \textbf{250185}  & \textbf{1.1200}  & \textbf{213.30} &  \textbf{736731}   &  \textbf{1.4166}  & \textbf{604.51} & \textbf{778269}   &  \textbf{1.5607}  & \textbf{652.30} \\  \hline
    
    \multirow{4}{*}{Trade-off} & Wallace\cite{Datapath-1964TC-Wallace} &  191214   &   1.1017   & 157.57 &  236566    &  1.2254  & 198.45 &   628322  &   1.6419  &518.51 &  735028  &  1.7352  & 601.46 \\ 
    &GOMIL\cite{Datapath-2021DATE-Xiao}      &  185357   &   1.0709   & 152.69 &  221857 & 1.2703  & 184.62 &  600947   &   1.5727  &496.70 & 649908  & 1.6847  & 555.77 \\ 
    &SA   &  169014    &  1.1079   & 139.35 & 204901 &    1.2552  &  181.19 &  580110      &  1.5959  &479.28 & 652019   & 1.7225  & 566.53 \\  
    &\textbf{RL-MUL}   &  167925   &  1.0263   & 137.64 & \textbf{203539}  &  \textbf{1.2353}  &  \textbf{179.80} &   578339    & 1.4987  &478.52  &  623691  &  1.6479  & 546.24 \\  
    &\textbf{RL-MUL 2.0}   &  \textbf{168606}   &  \textbf{0.9966}   & \textbf{138.06} & \textbf{203539}   &  \textbf{1.2353}  & \textbf{179.80} &  \textbf{576024}     & \textbf{1.4844}  & \textbf{475.76} &  \textbf{623555}   & \textbf{1.6479} & \textbf{545.83} \\
\hline
\end{tabular}
}
\label{table:pe-results}
\end{table}

\subsection{MAC Performance Comparison}
\begin{figure}
    \centering



\begin{tikzpicture}
\begin{groupplot}[group style={group size= 4 by 1,  horizontal sep=1.4cm, group name=myplot},height=3.4cm,width=3.8cm]
\nextgroupplot[minor tick num=0,
ymax=0.9,
xlabel={Area ($\upmu$m$^2$)},
ylabel={Delay (ns)},
y label style={at={(-0.2,0.5)}},
ymin=0.70,ymax=0.98,
ylabel near ticks,
legend style={
    draw=none,
	at={(0.5,1.15)},
	nodes={scale=0.75, transform shape},
	anchor=north,
	legend columns=-1,
}
]   
    \addplot[blue, style={mark=*, mark size=0.7pt, line width=0.8pt, draw=mydarkblue, draw opacity=0.8, mark options={fill opacity=0.8}}]  table [x=area, y=delay, col sep=comma] {pgfplot/data/mac_wallace_8.csv};
    \addplot[blue, style={mark=*, mark size=0.7pt, line width=0.8pt, draw=myblue, draw opacity=0.8, mark options={fill opacity=0.8}}]  table [x=area, y=delay, col sep=comma] {pgfplot/data/mac_ilp_8.csv};
    \addplot[blue, style={mark=*, mark size=0.7pt, line width=0.8pt, draw=myyellow, draw opacity=0.8, mark options={fill opacity=0.8}}]  table [x=area, y=delay, col sep=comma] {pgfplot/data/final_multi_constraint_sa_8bit_mac.csv};
    \addplot[blue, style={mark=*, mark size=0.7pt, line width=0.8pt, draw=myorange, draw opacity=0.8, mark options={fill opacity=0.8}}]  table [x=area, y=delay, col sep=comma] {pgfplot/data/final_multi_constraint_dqn_8bit_mac.csv};
    \addplot[blue, style={mark=*, mark size=0.7pt, line width=0.8pt, draw=mygreen, draw opacity=0.8, mark options={fill opacity=0.8}}]  table [x=area, y=delay, col sep=comma] {pgfplot/data/final_multi_constraint_a2c_8bit_mac.csv};
\coordinate (left) at (rel axis cs:0,1);

\nextgroupplot[minor tick num=0,
xlabel={Area ($\upmu$m$^2$)},
xtick={2000,2400},
ylabel={Delay (ns)},
y label style={at={(-0.2,0.5)}},
ymin=1.17,ymax=1.46,
ylabel near ticks,
legend style={
    draw=none,
	at={(1.0,1.35)},
	nodes={scale=0.75, transform shape},
	anchor=north,
	legend columns=-1,
}
]
    \addplot[blue, style={mark=*, mark size=0.7pt, line width=0.8pt, draw=mydarkblue, draw opacity=0.8, mark options={fill opacity=0.8}}]  table [x=area, y=delay, col sep=comma] {pgfplot/data/mac_wallace_16.csv};\addlegendentry{Wallace~\cite{Datapath-1964TC-Wallace}};
    \addplot[blue, style={mark=*, mark size=0.7pt, line width=0.8pt, draw=myblue, draw opacity=0.8, mark options={fill opacity=0.8}}]  table [x=area, y=delay, col sep=comma] {pgfplot/data/mac_ilp_16.csv};\addlegendentry{GOMIL~\cite{Datapath-2021DATE-Xiao}};
    \addplot[blue, style={mark=*, mark size=0.7pt, line width=0.8pt, draw=myyellow, draw opacity=0.8, mark options={fill opacity=0.8}}]  table [x=area, y=delay, col sep=comma] {pgfplot/data/final_multi_constraint_sa_16bit_mac.csv};\addlegendentry{SA};
    \addplot[blue, style={mark=*, mark size=0.7pt, line width=0.8pt, draw=myorange, draw opacity=0.8, mark options={fill opacity=0.8}}]  table [x=area, y=delay, col sep=comma] {pgfplot/data/final_multi_constraint_dqn_16bit_mac.csv};\addlegendentry{RL-MUL~\cite{DSE-2023DAC-Zuo}};
    \addplot[blue, style={mark=*, mark size=0.7pt, line width=0.8pt, draw=mygreen, draw opacity=0.8, mark options={fill opacity=0.8}}]  table [x=area, y=delay, col sep=comma] {pgfplot/data/final_multi_constraint_a2c_16bit_mac.csv};\addlegendentry{RL-MUL 2.0};
\coordinate (mid) at (rel axis cs:0.5,1);

\nextgroupplot[minor tick num=0,
xlabel={Area ($\cdot 10^5\upmu$m$^2$)},
ylabel={Delay (ns)},
y label style={at={(-0.2,0.5)}},
ylabel near ticks,
legend style={
    draw=none,
	at={(0.5,1.15)},
	nodes={scale=0.75, transform shape},
	anchor=north,
	legend columns=-1,
}
]
    \addplot[blue, style={mark=*, mark size=0.7pt, line width=0.8pt, draw=mydarkblue, draw opacity=0.8, mark options={fill opacity=0.8}}]  table [x=area, y=delay, col sep=comma] {pgfplot/data/pe_new/pe_mac_8bit_and_wallace_new.csv};
    \addplot[blue, style={mark=*, mark size=0.7pt, line width=0.8pt, draw=myblue, draw opacity=0.8, mark options={fill opacity=0.8}}]  table [x=area, y=delay, col sep=comma] {pgfplot/data/pe_new/pe_mac_8bit_and_ilp_new.csv};
    \addplot[blue, style={mark=*, mark size=0.7pt, line width=0.8pt, draw=myyellow, draw opacity=0.8, mark options={fill opacity=0.8}}]  table [x=area, y=delay, col sep=comma] {pgfplot/data/pe_new/pe_mac_8bit_and_sa_new.csv};
    \addplot[blue, style={mark=*, mark size=0.7pt, line width=0.8pt, draw=myorange, draw opacity=0.8, mark options={fill opacity=0.8}}]  table [x=area, y=delay, col sep=comma] {pgfplot/data/pe_new/pe_mac_8bit_and_dqn_new.csv};
    \addplot[blue, style={mark=*, mark size=0.7pt, line width=0.8pt, draw=mygreen, draw opacity=0.8, mark options={fill opacity=0.8}}]  table [x=area, y=delay, col sep=comma] {pgfplot/data/pe_new/pe_mac_8bit_and_a2c_new.csv};
\coordinate (right) at (rel axis cs:1,1);

\nextgroupplot[minor tick num=0,
xlabel={Area ($\cdot 10^5\upmu$m$^2$)},
ylabel={Delay (ns)},
y label style={at={(-0.2,0.5)}},
ymin=1.23,ymax=1.53,
ylabel near ticks,
legend style={
    draw=none,
	at={(1,1.35)},
	nodes={scale=0.75, transform shape},
	anchor=north,
	legend columns=-1,
}
]
    \addplot[blue, style={mark=*, mark size=0.7pt, line width=0.8pt, draw=mydarkblue, draw opacity=0.8, mark options={fill opacity=0.8}}]  table [x=area, y=delay, col sep=comma] {pgfplot/data/pe_new/pe_mac_16bit_and_wallace_new.csv};
    \addplot[blue, style={mark=*, mark size=0.7pt, line width=0.8pt, draw=myblue, draw opacity=0.8, mark options={fill opacity=0.8}}]  table [x=area, y=delay, col sep=comma] {pgfplot/data/pe_new/pe_mac_16bit_and_ilp_new.csv}; 
    \addplot[blue, style={mark=*, mark size=0.7pt, line width=0.8pt, draw=myyellow, draw opacity=0.8, mark options={fill opacity=0.8}}]  table [x=area, y=delay, col sep=comma] {pgfplot/data/pe_new/pe_mac_16bit_and_sa_new.csv};
    \addplot[blue, style={mark=*, mark size=0.7pt, line width=0.8pt, draw=myorange, draw opacity=0.8, mark options={fill opacity=0.8}}]  table [x=area, y=delay, col sep=comma] {pgfplot/data/pe_new/pe_mac_16bit_and_dqn_new.csv};
    \addplot[blue, style={mark=*, mark size=0.7pt, line width=0.8pt, draw=mygreen, draw opacity=0.8, mark options={fill opacity=0.8}}]  table [x=area, y=delay, col sep=comma] {pgfplot/data/pe_new/pe_mac_16bit_and_a2c_new.csv};
\coordinate (right) at (rel axis cs:1.5,1);

\end{groupplot}
\path (left)--(right) coordinate[midway] (group center);
\path (myplot c2r1.north west|-current bounding box.north)--
coordinate(legendpos)
(myplot c2r1.north west|-current bounding box.north);
\end{tikzpicture}
    \vspace{-.2in}
    \caption{Pareto-frontiers of the synthesis results on MACs and MAC-implemented PE arrays. From left to right: 8-bit MAC; 16-bit MAC; 8-bit MAC-implemented PE arrays; 16-bit MAC-implemented PE arrays.}
    \label{fig:results-mac}
\end{figure}
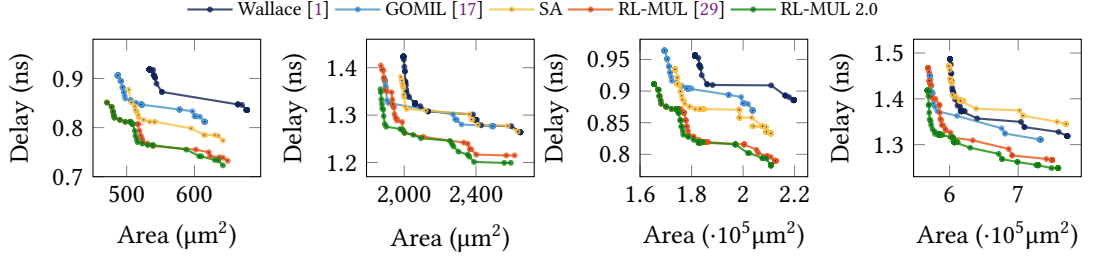

The curves in \Cref{fig:results-mac} for MACs and PE arrays consisting of MAC, along with the detailed comparisons in \Cref{table:mac-results}, demonstrate that RL-MUL 2.0 designs achieve superior performance compared to baselines. The RL-MUL 2.0 framework leads to up to a 13.4\% area reduction under the minimum area constraint and a 15.6\% decrease in delay under the minimum delay constraint for MACs. Similarly, PE arrays benefit from up to a 9.6\% reduction in area and a 13.1\% decrease in delay.

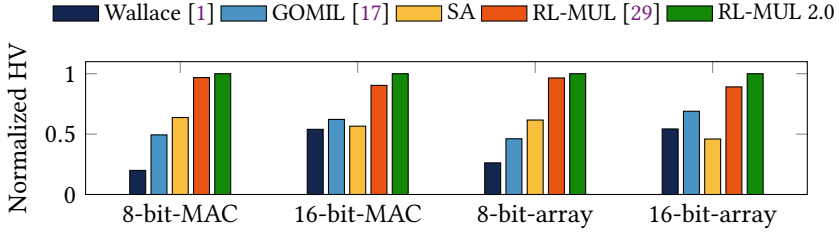
\begin{figure}[!tb]
    \centering
    \pgfplotsset{
	width =0.8\linewidth,
	height=0.24\linewidth
}

\begin{tikzpicture}
\begin{axis}[
    ybar,
    xticklabels={8-bit-MAC, 16-bit-MAC, 8-bit-array, 16-bit-array},xtick={0.8,2.3,3.8,5.3},
    xtick align=inside,
    ylabel={Normalized HV},
    ylabel near ticks,
    bar width = 6pt,
    xmin=0,
    xmax=6.1,
    ymin=0,
    ymax=1.1,
    legend style={at={(0.5,1.20)},
    draw=none,anchor=south,legend columns=-1,font=\small}
    ]

\addplot +[ybar, fill=mydarkblue,   draw=black, area legend] table [x={x},  y={wallace}] {hv_mac_new.dat};
\addplot +[ybar, fill=myblue,   draw=black, area legend] table [x={x},  y={ilp}] {hv_mac_new.dat};
\addplot +[ybar, fill=myyellow,   draw=black, area legend] table [x={x},  y={sa}] {hv_mac_new.dat};
\addplot +[ybar, fill=myorange,   draw=black, area legend] table [x={x},  y={dqn}] {hv_mac_new.dat};
\addplot +[ybar, fill=mygreen,   draw=black, area legend] table [x={x},  y={a2c}] {hv_mac_new.dat};
\legend{Wallace~\cite{Datapath-1964TC-Wallace}, GOMIL~\cite{Datapath-2021DATE-Xiao}, SA, RL-MUL~\cite{DSE-2023DAC-Zuo}, RL-MUL 2.0}
\end{axis}
\end{tikzpicture}
    \caption{Pareto-frontiers hypervolume comparison of MACs and MAC-implemented PE arrays.}
    \label{fig:hv_mac_pe}
\end{figure}

The hypervolume metrics, shown in \Cref{fig:hv_mac_pe} for MACs and PE arrays implemented by MAC, highlight RL-MUL 2.0's efficiency. RL-MUL 2.0 generates an average of 81.7\% more hypervolume than GOMIL for MACs and 80.9\% for the arrays. 
When comparing the performance of RL-MUL 2.0 to the original RL-MUL, there is a 7.0\% increase for MACs and a 7.9\% increase for arrays.

Regardless of the multiplier cases or the MAC cases, it is observed that the advantage of RL-MUL 2.0 over the SA approach varies between 8-bit and 16-bit configurations. GOMIL outperforms SA in larger bit widths, suggesting that evolutionary algorithms may struggle with large design spaces due to their complexity.
The improvement margins over the SA method vary between 8-bit and 16-bit designs, with GOMIL outperforming SA in larger bit widths. This suggests the evolutionary algorithm's limitations in addressing the expansive design space of larger bit widths. 
Additionally, the ILP-based method GOMIL simplifies the cost function by focusing solely on the area as the optimization objective. This approach limits its ability to achieve optimization gains in terms of delay consistently. In contrast, our approach employs multi-objective optimization, allowing us to attain Pareto-optimal results across both area and delay, demonstrating RL-MUL 2.0's consistent superiority across evaluations.



\begin{table}[!tb]
\caption{MUL and MAC area, timing, and power comparison (commercial synthesis tool).}
\centering
\resizebox{\linewidth}{!} {
\begin{tabular}{c|c|ccc|ccc|ccc|ccc}
    \hline
    \multirow{3}{*}{Preference} &\multirow{3}{*}{Method}& \multicolumn{6}{c|}{MUL} & \multicolumn{6}{c}{MAC} \\ \cline{3-14}
                      &      & \multicolumn{3}{c|}{8-bit} & \multicolumn{3}{c|}{16-bit} & \multicolumn{3}{c|}{8-bit} & \multicolumn{3}{c}{16-bit} \\ 
                        &    & Area ($\upmu$m$^2$) &  Delay (ns)& Power (uW) & Area ($\upmu$m$^2$) &  Delay (ns) & Power (mW) & Area ($\upmu$m$^2$) &  Delay (ns) & Power (uW) & Area ($\upmu$m$^2$) &  Delay (ns) & Power (mW) \\ \hline
    \multirow{4}{*}{Area} & Wallace\cite{Datapath-1964TC-Wallace}&  332.5000    &    1.5710    &  312.2091   &    1612.2260  &  2.4991    &  2.0378   &  442.6240  &  1.7983   &    473.6569  &  1846.3060    &  2.5995   &  2.4445  \\
    &GOMIL \cite{Datapath-2021DATE-Xiao}    &  326.4040    &  1.4703      &  313.3043    &  1570.2830   &  2.4098     &  1.9783    &  393.9460     &   1.6543    &  420.6916   &  1700.0060    &  2.5103    &  2.2788   \\ 
    &\textbf{RL-MUL}   &  322.6040     &   1.4578     &   196.7360    &   1550.5620    & 2.4004       &  1.9553    &  387.0640     &  1.6527    &   419.2456       &    1689.4040   &  2.5087     &  2.1896     \\  
    &\textbf{RL-MUL 2.0}   &  \textbf{320.5080}     &  \textbf{1.4367}      &  \textbf{196.3859}    &  \textbf{1538.0440}     &  \textbf{2.3857}     &  \textbf{1.9247}   &  \textbf{379.3560}     &  \textbf{1.6432}    &   \textbf{414.6432}    &   \textbf{1640.3560}   &  \textbf{2.4818}     &  \textbf{2.1752}     \\  \hline
    
    \multirow{4}{*}{Timing} & Wallace\cite{Datapath-1964TC-Wallace} &  464.1700     &  1.1124     &  399.9932    &    1904.5600   &  2.3413      &  2.2190    &  576.1560    &   1.2860   &  585.1385    &  2117.3600    &  2.4768    &  2.6243      \\
    &GOMIL \cite{Datapath-2021DATE-Xiao} &   440.7400    &   1.0789     &  378.6793    &   1770.2700     & 2.2812     & 2.1567      &   530.4040     &   1.2164    &  536.2910     &    1970.7940   & 2.4253     &   2.4654     \\  
    &\textbf{RL-MUL}   &  442.7080     &   1.0698     &   380.8763   &    1750.5080   &   2.2723    &   2.1367   &  531.8530     &  1.2167    &  533.7445    &   1972.6300   &  2.3464     &  2.4578    \\  
    &\textbf{RL-MUL 2.0}   &   \textbf{428.7080}    & \textbf{1.0567}      &  \textbf{366.7378}    &   \textbf{1744.7000}    &   \textbf{2.2660}    &  \textbf{2.1335}    &   \textbf{530.2460}    &  \textbf{1.2158}     &  \textbf{531.4654}     &  \textbf{1969.0400}    &  \textbf{2.339}     &  \textbf{2.4563}    \\  \hline
        \multirow{4}{*}{Trade-off} & Wallace\cite{Datapath-1964TC-Wallace} & 443.9540 & 1.1084 &388.6333  & 1776.8800 & 2.3172 & 2.1315 & 554.0780 & 1.3003 & 562.4348 & 1979.8380 & 2.5050 & 2.5145 \\
    &GOMIL \cite{Datapath-2021DATE-Xiao} &  431.4280 & 1.1024 &378.2456  & 1690.3600 & 2.2419 & 2.0643  & 514.1780 & 1.2153 & 531.8048  &1846.5720  & 2.4177 & 2.3804  \\  
    &\textbf{RL-MUL}   & 428.3760 & 1.0910 &372.3638 &  1678.0340 & 2.2406 & 2.0541 & 509.3560 & 1.2015 & 527.8372 &1816.3460  & 2.3935 & 2.3729  \\  
    &\textbf{RL-MUL 2.0}   &   \textbf{420.4060} & \textbf{1.0860} &\textbf{362.3837} &  \textbf{1669.4000} & \textbf{2.2387} & \textbf{2.0452}  & \textbf{504.5040} & \textbf{1.1893} & \textbf{521.3543} &  \textbf{1807.3040} & \textbf{2.3912} & \textbf{2.3679} \\  \hline
\end{tabular}
}
\label{table:dc-results}
\end{table}

\begin{table}[!tb]
\caption{MAC and PE array (MAC) area, timing, and power comparison.}
\vspace{-.1in}
\centering
\resizebox{\linewidth}{!} {
\begin{tabular}{c|c|ccc|ccc|ccc|ccc}
    \hline
    \multirow{3}{*}{Preference} &\multirow{3}{*}{Method}& \multicolumn{6}{c|}{MAC} & \multicolumn{6}{c}{PE-MAC} \\ \cline{3-14}
                      &      & \multicolumn{3}{c|}{8-bit} & \multicolumn{3}{c|}{16-bit} & \multicolumn{3}{c|}{8-bit} & \multicolumn{3}{c}{16-bit} \\ 
                        &    & Area ($\upmu$m$^2$) & Delay (ns) & Power (mW) & Area ($\upmu$m$^2$) & Delay (ns) & Power (mW) & Area ($\upmu$m$^2$) & Delay (ns) & Power (mW) & Area ($\upmu$m$^2$) & Delay (ns) & Power (mW) \\ \hline
    \multirow{4}{*}{Area} & Wallace\cite{Datapath-1964TC-Wallace}&  534    &   0.9182  & 0.5212 & 1995 & 1.4234 & 2.313 &  181340 & 0.9561 & 167.32 & 600471 & 1.4868 & 553.84 \\
    &GOMIL \cite{Datapath-2021DATE-Xiao}    &  487    &   0.9063    & 0.4674 & 1889  &  1.3787 & 2.167 &  169491 & 0.9638  & 156.65 &  571053   & 1.451 & 529.78 \\ 
    &SA   &  503   &  0.8775    &  0.4846& 1981  &   1.3807   &  2.289& 173577 & 0.9342 & 160.37 & 598836  & 1.4714 & 537.73 \\  
    &\textbf{RL-MUL}   &  \textbf{471}   &   \textbf{0.8511}   &  \textbf{0.4584}& 1870  &  1.4046    &  2.169&  165405 & 0.9112  & 154.41 &  568465 &  1.4673 & 523.36 \\  
    &\textbf{RL-MUL 2.0}   &   \textbf{471}  &    \textbf{0.8511}  &  \textbf{0.4584}& \textbf{1868} &   \textbf{1.3545}   &  \textbf{2.158}&  \textbf{165405} & \textbf{0.9112} & \textbf{154.38} &  \textbf{567784} & \textbf{1.4178}  & \textbf{522.58}  \\  \hline
    
    \multirow{4}{*}{Timing} & Wallace\cite{Datapath-1964TC-Wallace} &  677    &   0.8359  & 0.7392 & 2646 & 1.264    & 3.291 &  219678 & 0.8856 & 202.61 & 771664 & 1.3187 & 710.85 \\
    &GOMIL \cite{Datapath-2021DATE-Xiao} & 615     &  0.8119    & 0.6472 &  2494    &  1.2766& 3.039 & 203743 &   0.8693 & 188.57 & 730670  & 1.3109 & 675.11 \\ 
    &SA   &  642   &   0.7737   &  0.6865& 2632  &  1.2652    &  3.282& 210825 & 0.8331 & 195.18 & 769893  &  1.3448 & 694.90 \\  
    &\textbf{RL-MUL}   &  649   &   0.7324   &  0.6983& 2568  &   1.2149   &  3.110&   212596   & 0.7897 & 197.23 &  749533  & 1.2668  & 677.48 \\  
    &\textbf{RL-MUL 2.0}   &  \textbf{642}   &   \textbf{0.7231}   &  \textbf{0.6948}& \textbf{2594}  &    \textbf{1.1992}  &  \textbf{3.192}&   \textbf{210825} & \textbf{0.7827} & \textbf{195.18} &  \textbf{758385} &   \textbf{1.2487} & \textbf{683.39}\\  \hline

    \multirow{4}{*}{Trade-off} & Wallace \cite{Datapath-1964TC-Wallace}&  552   &   0.8727 & 0.5450 & 2060 &   1.3248   & 2.432 & 186038 & 0.9107 & 173.17 & 611502 & 1.3851 & 563.23  \\    
    &GOMIL\cite{Datapath-2021DATE-Xiao}  &498  &  0.859 & 0.4857 &  2486 &  1.2766 & 3.014 &  172283 &  0.9161  & 160.37 &  582085 & 1.3729 & 535.26 \\ 
    &SA   &  518   &   0.8202   &  0.5043& 2005  &   1.3181   &  2.335& 177322 & 0.8799 & 166.99 & 604148  & 1.4084 & 545.24 \\  
    &\textbf{RL-MUL}   &   \textbf{482}  &   \textbf{0.8202}   &  \textbf{0.4774}& 1946  &   1.3016   &  2.329& 168197 & 0.8799 & 158.36 & 588486 & 1.3429  & 529.68 \\  
    &\textbf{RL-MUL 2.0}   &  \textbf{482}   &  \textbf{0.8202}    &  \textbf{0.4774}&  \textbf{2002} &  \textbf{1.2625}    &  \textbf{2.350}&  \textbf{176777} & \textbf{0.8309} & \textbf{166.89} &  \textbf{578816} &  \textbf{1.3233} & \textbf{529.38} \\ 
    \hline
\end{tabular}
}
\label{table:mac-results}
\end{table}

\subsection{Efficient and Stable Training }
\label{subsec:exp-efficient&stable}
\begin{figure}[!tb]
    \centering
    \includegraphics[width=0.9\linewidth]{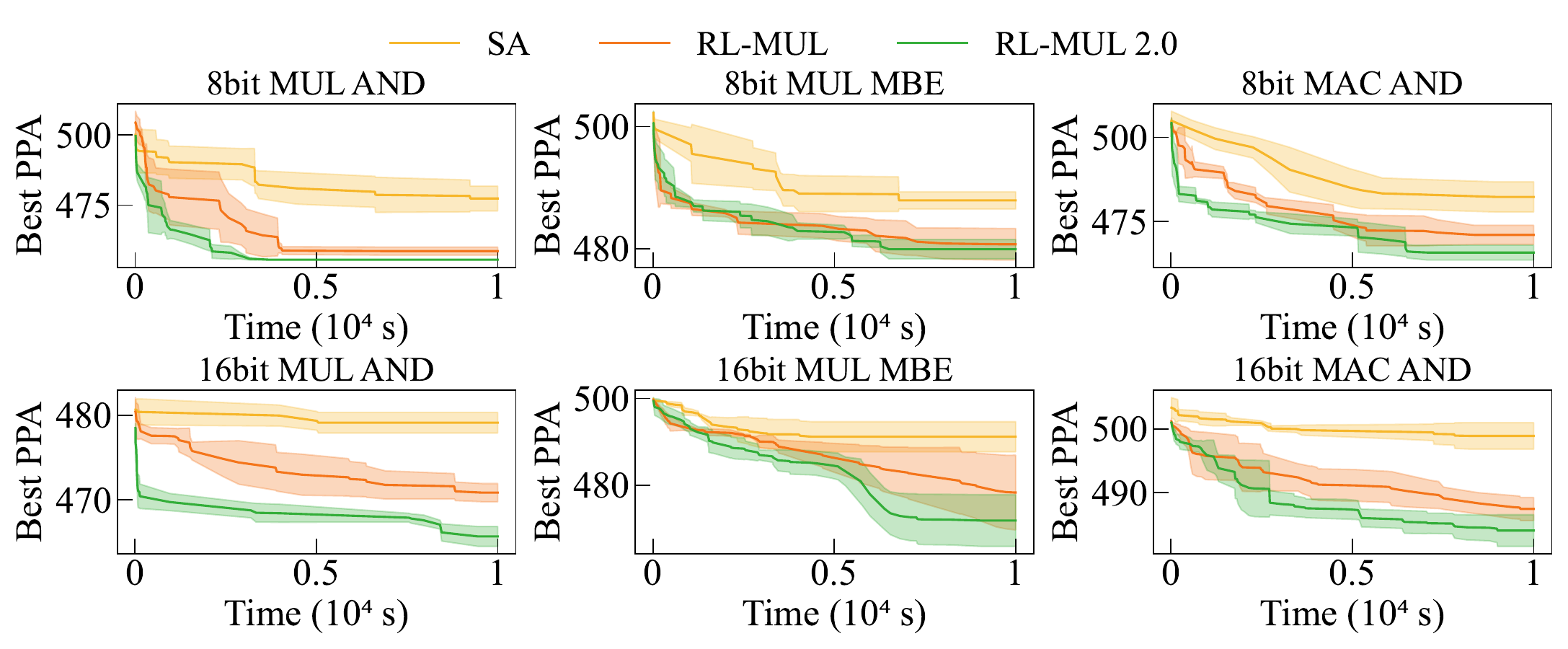} 
    \caption{Optimization trajectories for different methods illustrate the mean PPA $\pm $ standard error. The shaded areas represent the standard deviation of the PPA values.}
    \label{fig:trajectories}
\end{figure}

We conducted six experiments, each repeated three times, on the original RL-MUL, RL-MUL 2.0, and SA algorithms, with a fixed PPA weight across two bit-widths. 
These experiments are categorized into three groups: one focusing on AND-based MUL operations, another on MUL operations employing Booth encoding, and a third on MAC operations. 
The mean PPA values are represented by a solid line, with the standard deviation depicted as the surrounding shadow in \Cref{fig:trajectories}.
Across all datasets, our RL methods consistently demonstrate superior performance, significantly outperforming SA. Particularly, the RL-MUL 2.0 demonstrates superior results and a more stable and efficient training process.
Furthermore, it is observed that there exists a gap between the variance shadow caused by independent repetitions of RL-MUL and RL-MUL 2.0 experiments, especially in the 16-bit designs. 
So it implies that even if the DQN in RL-MUL were to support parallel agents within the 10,000s runtime, the best results achieved are still not as good as RL-MUL 2.0.

When comparing efficiency, runtime serves as a key metric, reflecting the algorithm's ability to explore the design space within a given time. Parallel processing allows RL-MUL 2.0 to utilize computational power to accelerate exploration. 
Notably, for the 8-bit MUL AND case, RL-MUL 2.0 achieves the optimal PPA value reached by RL-MUL in an average of 2,124 seconds. Similarly, RL-MUL 2.0 reaches this level in 6,275 seconds for the 8-bit MUL MBE case, 5,166 seconds for the 8-bit MAC AND case, 136 seconds for the 16-bit MUL AND case, 5,985 seconds for the 16-bit MUL MBE case, and 4,823 seconds for the 16-bit MAC AND case. These results indicate that RL-MUL 2.0 requires significantly less time to achieve the same performance level as RL-MUL, demonstrating the efficiency of its parallel training approach.

In addition to OpenROAD flow and OpenSTA, we conducted a cross-check synthesis using Synopsys Design Compiler \cite{DesignCompiler} to validate the multipliers and MACs from our RL-MUL framework. 
The results, shown in \Cref{table:dc-results}, demonstrate that RL-MUL-designed multipliers consistently outperform baseline multipliers, achieving superior area, delay, and power metrics under commercial EDA tools, thereby confirming the effectiveness and robustness of our designs across synthesis environments.

\section{Conclusion}
\label{sec:conclu}
In this research, we introduce a novel framework for optimizing multipliers through reinforcement learning. 
The framework utilizes an RL agent that adapts based on EDA tool feedback to engineer multipliers achieving Pareto optimality. 
We demonstrate that multipliers and MACs designed by RL can Pareto-dominate multipliers that are produced by existing approaches.
The obtained optimized multiplier and MACs can be further applied in the implementation of a larger module, such as a PE array. 
Looking ahead, we aim to broaden the application of our RL methodology to encompass more extensive datapath components, enhancing the scope and impact of our optimization efforts.
{
    \bibliographystyle{IEEEtran}
    \bibliography{ref/Top, ref/RL}
}

\end{document}